\documentclass[english,notitlepage,nofootinbib]{revtex4-1}
\pdfoutput=1
\usepackage[T1]{fontenc}
\usepackage[latin9]{inputenc}
\usepackage{xcolor}
\usepackage{amsmath}
\usepackage{amssymb}
\usepackage{graphicx}
\newcommand{\boss}[2]{\ensuremath{\rlap{\kern-2.5pt\ensuremath{\overset{\scriptscriptstyle(-)}{\phantom{#1}}}}{\ensuremath{{#1}_{#2}}}}}

\makeatletter

\providecommand{\tabularnewline}{\\}

\makeatother

\usepackage{babel}
\begin{document}
\title{Searching for BSM neutrino interactions in  dark matter detectors}
\author{Jonathan M. Link $^a$, and Xun-Jie Xu $^b$}
\affiliation{$^{a}$ Center for Neutrino Physics, Physics Department, Virginia
Tech, 850 West Campus Dr, Blacksburg, VA 24061, USA.\textcolor{black}{}~\\
\textcolor{black}{$^{b}$ Max-Planck-Institut f\"ur Kernphysik, Postfach
103980, D-69029 Heidelberg, Germany}.}
\date{\today}
\begin{abstract}
Neutrino interactions beyond the Standard Model (BSM) are theoretically
well motivated and have an important impact on the future precision measurement
of neutrino oscillation. In this work, we study the sensitivity of
a multi-ton-scale liquid Xenon dark matter detector equipped with
an intense radioactive neutrino source to various BSM neutrino-electron
interactions. We consider the conventional Non-Standard Interactions
(NSIs), other more generalized four-fermion interactions including
scalar and tensor forms, and light-boson mediated interactions. The
work shows that with realistic experimental setups, one can achieve
unprecedented sensitivity to these BSM neutrino-electron interactions.
\end{abstract}
\maketitle

\section{Introduction}

The existence of small neutrino masses is calling for new physics.
It is fairly reasonable to speculate that the interactions of neutrinos
may also go beyond the Standard Model (BSM). Searching for BSM neutrino
interactions is of increasing importance since they could make a considerable
impact on future precision oscillation measurements.  For example, it is known that the Non-Standard Interactions~\cite{Davidson:2003ha,Ohlsson:2012kf,Farzan:2017xzy,Esteban:2018ppq}
of neutrinos may affect the measurement of $\delta_{CP}$ in long-baseline
oscillation experiments~\cite{Forero:2016cmb,Miranda:2016wdr,Masud:2015xva,Bakhti:2016prn,Masud:2016bvp,C.:2017yqh,Hyde:2018tqt,Deepthi:2017gxg}. Moreover, due to various well-known problems of the SM, 
theorists have for decades been looking for more satisfactory theories. As neutrino
masses and mixing have become established, precision measurement of neutrino 
interactions are increasingly important in directing theoretical exploration. 

Actually, if BSM neutrino interactions exist, charged leptons such
as the electron are  very likely to participate in these interactions
as well, since neutrinos and charged leptons are embedded in $SU(2)_{L}$
doublets in the SM, and new interactions are generally expected to
couple with the whole doublet.  Many BSM theories such as Type II seesaw
models~\cite{Konetschny:1977bn,Cheng:1980qt,Lazarides:1980nt,Mohapatra:1980yp,Schechter:1980gr,Schechter:1981cv},
left-right symmetric models~\cite{Pati:1974yy,Mohapatra:1974gc,Senjanovic:1975rk},
and the various $U(1)$ extensions~\cite{Jenkins:1987ue,He:1990pn,Buchmuller:1991ce,He:1991qd,Foot:1994vd,Emam:2007dy,Basso:2008iv},
are some typical examples which introduce new interactions between neutrinos and 
charged leptons. 

Therefore, searching for neutrino-electron BSM interactions is theoretically
highly motivated and has been extensively studied in the literature~\cite{Sevda:2016otj,Deniz:2009mu,Forero:2011zz,Kaneta:2016uyt,Rodejohann:2017vup,Lindner:2018kjo,Bischer:2018zcz,Arguelles:2018mtc,Deniz:2017zok,Kouzakov:2017hbc,Khan:2017oxw,Khan:2016uon,Khan:2017djo,Babu:2017olk,Campos:2017dgc,Bauer:2018onh}.
So far  the most precise measurement for $\nu_{\mu}$($\overline{\nu}_{\mu}$)-$e$ elastic 
scattering\footnote{Quasi-elastic scattering processes such as 
$\nu_{\mu(\tau)}+e\rightarrow\mu(\tau)+\nu_{e}$ have little potential of 
finding new physics since the corresponding charged current interactions have 
been very precisely determined by $\mu$ and $\tau$ decays.}, comes from the 
CHARM II experiment~\cite{Vilain:1993kd,Vilain:1994qy}, which achieved a 
precision of 3\%. For $\nu_{e}$-$e$ and $\overline{\nu}_{e}$-$e$
scattering, the best measurements come from LSND \cite{Auerbach:2001wg}
and TEXONO~\cite{Deniz:2009mu} respectively, with only 20\% precision.
Recently there have been proposals~\cite{Pospelov:2011ha,Harnik:2012ni,Pospelov:2012gm,Pospelov:2013rha,Coloma:2014hka}
to using dark matter detectors with intense radioactive neutrino sources
to measure neutrino scattering, that could lead to significant improvements.  
In particular, ton-scale liquid xenon (LXe) detectors~\cite{Malling:2011va,Akerib:2015cja,Mount:2017qzi}
equipped with state-of-the-art $^{51}{\rm Cr}$ sources~\cite{Cribier:1996cq}
(referred to as $^{51}$Cr-LXe throughout this paper), with their 
combination of a large mono-energetic $\nu_e$ flux, high electron density, 
and low backgrounds, offer a path to reaching unprecedented precision. 

In this paper, we study the potential of $^{51}$Cr-LXe
experiments for probing BSM neutrino interactions.   Our previous
study~\cite{Coloma:2014hka} showed that the combination of a 5~MCi 
$^{51}{\rm Cr}$ source and a 6~ton LXe detector could generate the 
tightest terrestrial constraint on neutrino magnetic moments, and competitive
sensitivity to sterile neutrino oscillation.  Therefore, we expect
similar setups would have excellent sensitivity to a wide range of BSM 
neutrino interactions, including Non-Standard Interactions (NSIs), general
four-fermion effective interactions (e.g., including tensor and scalar
forms), and light and weakly coupled mediators (e.g. dark photons).
The new physics scenarios considered in this work can also be tested in some other low-energy processes such as coherent elastic neutrino-nucleus  scattering (CE$\nu$NS) and atomic parity violation (APV)---for a recent study, see Ref.~\cite{Arcadi:2019uif}. The main difference is that CE$\nu$NS and APV test interactions between leptons (neutrinos or electrons) and quarks, while $\nu$-$e$ scattering experiments test purely leptonic interactions between neutrinos and electrons. Although phenomenologically they are different, theoretically these interactions would be very closely correlated in some models. So combining both types of low-energy measurements could provide a complementary test of BSM theories.

The paper is organized as follows. In Sec.~\ref{sec:basic}, we introduce
the basic experimental setup and evaluate the event rates of signals
and backgrounds for several configurations. In Sec.~\ref{sec:SM},
we study the precision measurement of the SM parameters relevant to
neutrino interactions. In Sec.~\ref{sec:BSM}, we include sensitivity 
studies for several types of BSM neutrino interactions which are widely 
discussed in the literature.  Finally, we conclude and summarize
the results in Sec.~\ref{sec:Conclusion}.

\section{Event rates\label{sec:basic}}
Let us consider  a general $^{51}$Cr-LXe experiment and evaluate
the event numbers of $\nu$-$e$ scattering. In the $^{51}$Cr-LXe
experiment, the number of $\nu$-$e$ scattering events appearing in the 
infinitesimal volume $dV$ around position $\vec{r}\equiv(x,\ y,\ z)$ in 
the detector, during the time interval $t$ to $t+dt$, with the recoil 
energy from $T$ to $T+dT$ can be evaluated as 
\begin{equation}
dN=dV\, n_{e}\, dt\, dT\left[\int\phi\left(\vec{r},\ t,\ E_{\nu}\right)\frac{d\sigma}{dT}\left(T,\ E_{\nu}\right)dE_{\nu}\right],\label{eq:LXe-4}
\end{equation}
where $n_{e}$, $\phi$, $\frac{d\sigma}{dT}$, and $E_{\nu}$ denote
the electron number density, the neutrino flux, the differential cross
section, and the neutrino energy respectively.

For a point-like radioactive source, $\phi$ has the following spatial
and temporal dependence:
\begin{equation}
\phi\left(\vec{r},\ t,\ E_{\nu}\right)=\phi_{0}R_{{\rm Cr51}}^{0}\frac{1\,{\rm m}^{2}}{r^{2}}e^{-t/\tau}f\left(E_{\nu}\right),\label{eq:LXe-5}
\end{equation}
where
\begin{equation}
\phi_{0}=2.94\times10^{15}\ {\rm neutrinos}/({\rm MCi\ m^2\,s}),\label{eq:LXe-6}
\end{equation}
is the neutrino flux at 1 meter
from a 1~MCi radioactive source; $\tau=39.96$ days is the mean 
lifetime of $^{51}{\rm Cr}$; $R_{{\rm Cr51}}^{0}$ is the initial ($t=0$) radioactivity; and  $f(E_{\nu})$ describes the spectral shape, normalized by $\int f(E_{\nu}) dE_{\nu}=1$.  
The neutrino spectrum of $^{51}{\rm Cr}$ simply consists of mono-energetic
neutrino emissions at 750 keV (90\%) and 430 keV (10\%):
\begin{equation}
f(E_{\nu})=0.9\,\delta(E_{\nu}\!-\!750\,{\rm keV})+0.1\,\delta(E_{\nu}\!-\!430\,{\rm keV}). \label{eq:f}
\end{equation}

Since the source decays exponentially, it is 
useful to define a time-averaged activity,
\begin{equation}
\langle R_{{\rm Cr51}}\rangle\equiv\frac{R_{{\rm Cr51}}^{0}}{\Delta t}\int_{0}^{\Delta t}dte^{-t/\tau}=\frac{\tau R_{{\rm Cr51}}^{0}}{\Delta t}\left[1-e^{-\Delta t/\tau}\right],\label{eq:phi_avg}
\end{equation}
relative to the initial activity, $R_{{\rm Cr51}}^{0}$, and the 
exposure time, $\Delta t$.  Likewise, an average distance between 
the detector and the source, $r_{{\rm avg}}$, can be defined as:
\begin{equation}
\frac{1}{r_{{\rm avg}}^{2}}\equiv\langle\frac{1}{r^{2}}\rangle=\frac{1}{V}\int\frac{dV}{r^{2}}.\label{eq:r_eff}
\end{equation}
With these averaged values, it is equivalent use the following replacement 
in our analyses:
\begin{equation}
\phi\left(\vec{r},\ t,\ E_{\nu}\right) \rightarrow \phi_0\frac{1\,{\rm m^2}}{r_{{\rm avg}}^{2}} \langle R_{{\rm Cr51}}\rangle f\left(E_{\nu}\right) \equiv \phi_{\rm avg}\left(E_{\nu}\right), \label{eq:ph_avg}
\end{equation}
while collapsing the time and space integrals to the $\Delta t$ and 
the detector fiducial volume, $V$.  Hence the number of events in the 
recoil energy bin $[T_{i},\ T_{i}+\Delta T]$ can be written as 
\begin{equation}
N_{i}=Vn_{e}\,\Delta t\,\int_{T_{i}}^{T_{i}+\Delta T}\Phi(T)dT,\label{eq:LXe}
\end{equation}
with
\begin{equation}
\Phi(T)\equiv\int\phi_{{\rm avg}}\left(E_{\nu}\right)\frac{d\sigma}{dT}\left(T,\ E_{\nu}\right)dE_{\nu},\label{eq:LXe-1}
\end{equation}
where $\phi_{{\rm avg}}\left(E_{\nu}\right)$ has been defined in Eq.~(\ref{eq:ph_avg}).

\begin{table*}[b]
\begin{ruledtabular}
\begin{tabular}{ccccccc}
 &  $R_{{\rm Cr51}}^{0}$ & $\Delta t$ & $\langle R_{{\rm Cr51}}\rangle$  & background 
 & $r_{{\rm avg}}$
 & $Vn_{e}$\tabularnewline
\hline 
Configuration A & 5 MCi $^{51}{\rm Cr}$ & 100 days & 1.83 MCi & normal  & $1.63\ {\rm m}$ & $1.5\times10^{30}$\tabularnewline
Configuration B & 5 MCi $^{51}{\rm Cr}$ & 50 days & 2.85 MCi & $^{136}{\rm Xe}$ depleted & $1.63\ {\rm m}$ & $1.5\times10^{30}$\tabularnewline
Configuration C & 10 MCi $^{51}{\rm Cr}$ & 50 days & 5.71 MCi & $^{136}{\rm Xe}$ depleted  & $1.63\ {\rm m}$ & $1.5\times10^{30}$\tabularnewline
\end{tabular}
\end{ruledtabular}
\vspace{-3mm}
\caption{\label{tab:t}Configurations of the $^{51}$Cr-LXe experiment
considered in this work.}
\end{table*}


For a detector with a cylindrical fiducial region, $r_{{\rm avg}}$
can be computed analytically (see Appendix~\ref{sec:Geometrical}).
We take the same geometrical profile as Ref.~\cite{Coloma:2014hka},
i.e., the height $(h)$ and diameter ($d$) of the cylinder are $h=d=1.38$
m, and the source is $L=1$ m below the bottom. Using Eq.~(\ref{eq:r_eff_analytic}),
this profile has $r_{{\rm avg}}=1.63$ m and contains 6 tons of LXe,
which corresponds to $Vn_{e}=$$1.5\times10^{30}$~electrons. 

Plugging in the definition of $\phi_{\rm avg}$, the integral in Eq.~(\ref{eq:LXe-1})
works out as
\begin{equation}
\Phi(T)=\frac{1\,{\rm m^2}}{r_{{\rm avg}}^{2}} \langle R_{{\rm Cr51}}\rangle \phi_0\left[0.9\,\frac{d\sigma}{dT}(T,\ E_{\nu}\!=\!750\,{\rm keV})+0.1\,\frac{d\sigma}{dT}(T,\ E_{\nu}\!=\!430\,{\rm keV})\right].\label{eq:LXe-3}
\end{equation}
In practical calculations, we also need the maximal recoil energy, $T_{\max}$, which is defined as the maximal recoil of electron that can be generated by a certain $E_{\nu}$: 
\begin{equation}
T_{\max}=\frac{2E_{\nu}^{2}}{2E_{\nu}^{2}+m_{e}}\approx\begin{cases}
559\ {\rm keV} & (E_{\nu}=750\,{\rm keV})\\
270\ {\rm keV} & (E_{\nu}=430\,{\rm keV})
\end{cases}.\label{eq:LXe-33}
\end{equation}
For $T>T_{\max}$ the corresponding $\frac{d\sigma}{dT}$ in Eq.~(\ref{eq:LXe-3}) is set to zero. 

\begin{figure}[b]
\centering

\includegraphics[width=12cm]{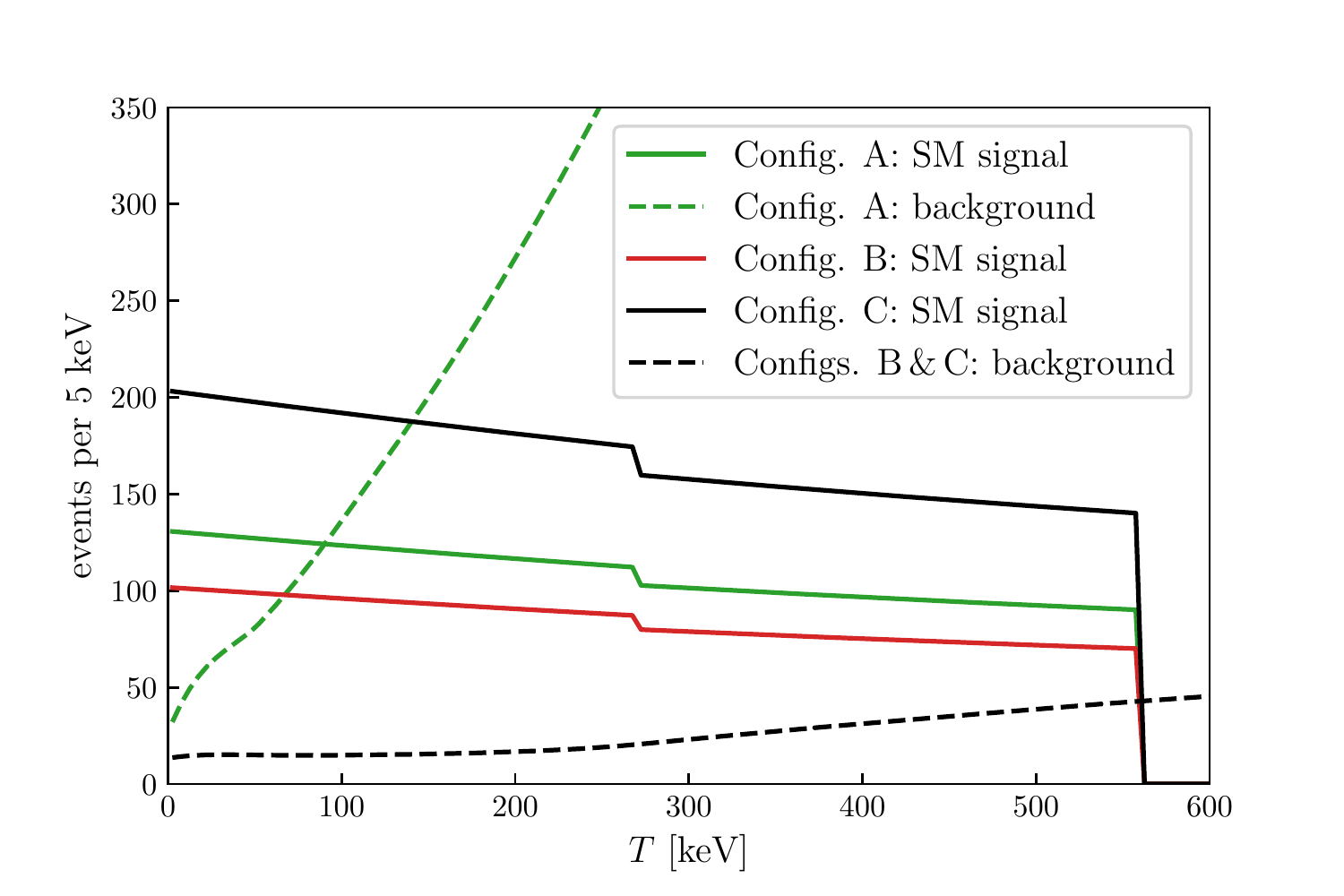}

\caption{\label{fig:Signal}Signal compared to background in the $^{51}$Cr-LXe
experiment.  Three configurations (A, B, C) are considered, with the
details described in Tab.~\ref{tab:t}. Configurations B and C have
much larger signal/background ratios than A. }
\end{figure}

In Tab.~\ref{tab:t}, we summarize useful quantities for the
configurations considered in this work. The event numbers with the SM
$\nu$-$e$ elastic scatting cross section~\cite{Giunti} are 
evaluated using Eq.~(\ref{eq:LXe}) and presented in Fig.~\ref{fig:Signal}.
Regarding backgrounds, we assume a ``normal'' level~\cite{Baudis:2013qla},
which includes the solar neutrino background and decays of embedded 
radioactive isotopes ($^{136}{\rm Xe}$, $^{85}{\rm Kr}$, $^{222}{\rm Rn}$, 
etc.), for configuration A; and a ``$^{136}{\rm Xe}$ depleted'' level, which 
consists of the normal solar background with only 10\% of the normal level 
from radioactive decays, for configurations B and C\@.  In configurations 
B and C, the background rate is further decreased by halving the exposure 
time, $\Delta t$.  Finally, in configuration C, the signal-to-noise ratio 
is further increased by the use of a more intense source.

\section{Measurement of the SM parameters\label{sec:SM}}
The study of neutrino-electron scattering has played an important role in the 
precision electroweak tests of the SM\,---\,see Chapter 10 of PDG \cite{Tanabashi:2018oca}.
For instance, the CHARM II experiment, which has performed the most 
precise measurement of $\nu_{\mu}$-$e$ and $\overline{\nu}_{\mu}$-$e$
scattering thus far, has determined the electroweak mixing angle, 
$\sin^{2}\theta_{W}$, to be:
\begin{equation}
\sin^{2}\theta_{W}=0.2324\pm0.0083,\ \ ({\rm CHARM\ II,\ 1994}),\label{eq:CHARMII}
\end{equation}
with about 3\% precision~\cite{Vilain:1994qy}.  In comparison $\boss{\nu}{e}$ 
scattering experiments have not yet achieved such precision.  The best 
measurements of $\nu_{e}$-$e$ and $\overline{\nu}_{e}$-$e$ scattering 
come from LSND \cite{Auerbach:2001wg} and TEXONO \cite{Deniz:2009mu}
respectively:
\begin{equation}
\sin^{2}\theta_{W}=\begin{cases}
0.248\pm0.051 & (\nu_{e}$-$e\ {\rm LSND})\\
0.251\pm0.031\pm0.024 & (\overline{\nu}_{e}$-$e\ {\rm TEXONO})
\end{cases}.\label{eq:LSND-TEXONO}
\end{equation}
We will show that a $^{51}$Cr-LXe experiment could provide
a superior precision measurement of $\sin^{2}\theta_{W}$, which 
not only exceeds that of these aforementioned experiments, but also 
reverses the current situation in which $\boss{\nu}{e}$ scattering is 
less precise than $\boss{\nu}{\mu}$ scattering. 

In the SM, the low energy interactions of neutrinos and electrons
can be  described by the effective Lagrangian 
\begin{equation}
{\cal L}_{{\rm NC}+{\rm CC}}\supset\frac{G_{F}}{\sqrt{2}}\overline{\nu}_{e}\gamma^{\mu}(1-\gamma^{5})\nu_{e}\overline{\psi_{e}}\gamma^{\mu}\left[g_{V}-g_{A}\gamma^{5}\right]\psi_{e},\label{eq:LXe-9}
\end{equation}
where, for $\nu_e$-$e$ scattering, $g_{V}^e$ and $g_{A}^e$ are given at the tree level by
\begin{equation}
g_{V}^e=2\sin^{2}\theta_{W}+\frac{1}{2}\ \ {\rm and}\ \ g_{A}^e=\frac{1}{2},\label{eq:gVgA}
\end{equation}
which includes both the neutral current (NC) and the charged current
(CC) contributions. 
At the tree level, the differential cross section for $\nu$-$e$
scattering is given by~\cite{Giunti}:
\begin{equation}
\frac{d\sigma}{dT}=\frac{m_{e}G_{F}^{2}}{2\pi}\left[g_{1}^2+g_{2}^{2}\left(1-\frac{T}{E_{\nu}}\right)^{2}-g_{1}g_{2}\frac{m_{e}T}{E_{\nu}^{2}}\right],\label{eq:LXe-7}
\end{equation}
where $T$ is the electron's recoil energy and 
\begin{equation}
    g_1\equiv g_V+g_A\ \ {\rm and}\ \ g_2\equiv g_V-g_A.\label{eq:LXe-8}
\end{equation}

The experimental value of $\sin^{2}\theta_{W}$ is well determined at the $Z$-pole and found to be $0.23122$ (defined in the Modified Minimal Subtraction ($\overline{{\rm MS}}$) scheme~\cite{Tanabashi:2018oca}).
At low energies ($\lesssim0.1$ GeV), its value can be extrapolated theoretically using renormalization group equation (RGE) running \cite{Erler:2004in}:
\begin{equation}
\text{\ensuremath{\sin^{2}\theta_{W}}\ensuremath{\ensuremath{\approx}}}\,0.23865\ \ \ ({\rm low\ energy\ limit}).\label{eq:LXe-11}
\end{equation}
Including the radiative corrections, the expressions of $g_{V}$ and $g_{A}$
should also be modified \cite{Erler:2013xha}: 
\begin{equation}
g_{V}^e\approx 2.00128\sin^{2}\theta_{W}\!+\!0.4828\ \ {\rm and}\ \ g_{A}^e\approx 0.4936.\label{eq:LXe-17}
\end{equation}

\begin{figure}
\centering

\includegraphics[width=12cm]{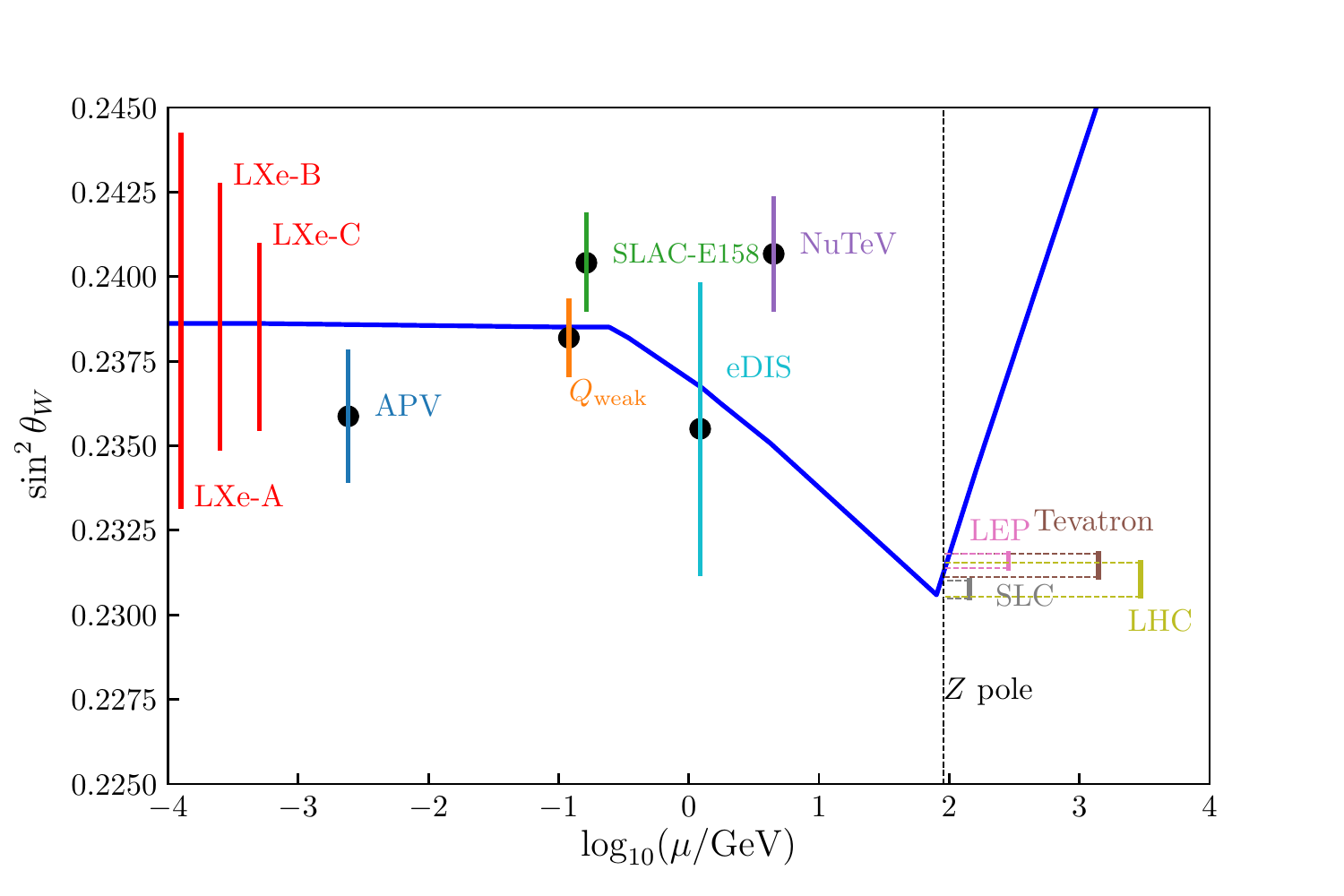}

\caption{\label{fig:sw}Measurement of $\sin^2\theta_{W}$ at different energy scales
($\mu$). The red error bars represent the precision of LXe-A, LXe-B, and LXe-C 
expected from this work.  The other measurements are from Fig.~10.2 of the 
PDG~\cite{Tanabashi:2018oca}. Measurements from earlier neutrino elastic 
scattering experiments do not have sufficient precision to be included in this 
plot.}
\end{figure}
Taking $\sin^{2}\theta_{W}$ as a varying parameter, one can measure
it in the $^{51}$Cr-LXe experiment, as a low-energy
precision test of the SM\@.  The simulated data is generated assuming
the numerical values in Eqs.~(\ref{eq:LXe-11}) and (\ref{eq:LXe-17}),
and including the backgrounds shown in Fig.~\ref{fig:Signal}. Then
we perform a $\chi^{2}$-fit on the data to obtain the 1$\sigma$
confidence level (CL) of $\sin^{2}\theta_{W}$.  The results are 
presented in Fig.~\ref{fig:sw} and Tab.~\ref{tab:sw}. 
\begin{table*}[b]
\begin{ruledtabular}
\begin{tabular}{cccc}
 & Configuration A & Configuration B & Configuration C\tabularnewline
\hline 
precision of $\sin^{2}\theta_{W}$ & $0.23865\pm0.0055$ & $0.23865_{-0.0037}^{+0.0041}$ & $0.23865_{-0.0031}^{+0.0023}$\tabularnewline
relative precision & $\pm2.3\%$ & $_{-1.6\%}^{+1.7\%}$ & $_{-1.3\%}^{+1.0\%}$\tabularnewline
\end{tabular}
\end{ruledtabular}
\vspace{-3mm}
\caption{\label{tab:sw}Precision of the $\sin^{2}\theta_{W}$ measurement in the three $^{51}$Cr-LXe configurations considered.}
\end{table*}
Comparing these results with the current state-of-the-art from CHARM II, LSND,
and TEXONO, we see that the $^{51}$Cr-LXe experiment is able to reach
unprecedented precision on $\sin^{2}\theta_{W}$ as measured by $\nu$-$e$
scattering, even in the least sensitive configuration that we have considered.

In Fig.~\ref{fig:sw}, we show the role of $^{51}$Cr-LXe measurements
in probing the RGE running of $\sin^{2}\theta_{W}$. So far, the most
precise measurements of $\sin^{2}\theta_{W}$ are from colliders,
including LEP, LHC, SLC, and Tevatron.  All the colliders essentially
measure the value of $\sin^{2}\theta_{W}$ only at the $Z$-pole scale.  Other measurements,
including the weak charge of protons ($Q_{{\rm weak}}$), Atomic Parity
Violation (APV), and electron deep-inelastic scattering (eDIS), are
located at various lower energy scales. This plot shows that a precision
probe of the RGE running of $\sin^{2}\theta_{W}$, is still needed at the 
lowest energies, which a $^{51}$Cr-LXe measurement (marked in red as LXe-A, B 
and C) would provide.

\begin{figure}[b]
\centering
\includegraphics[width=12cm]{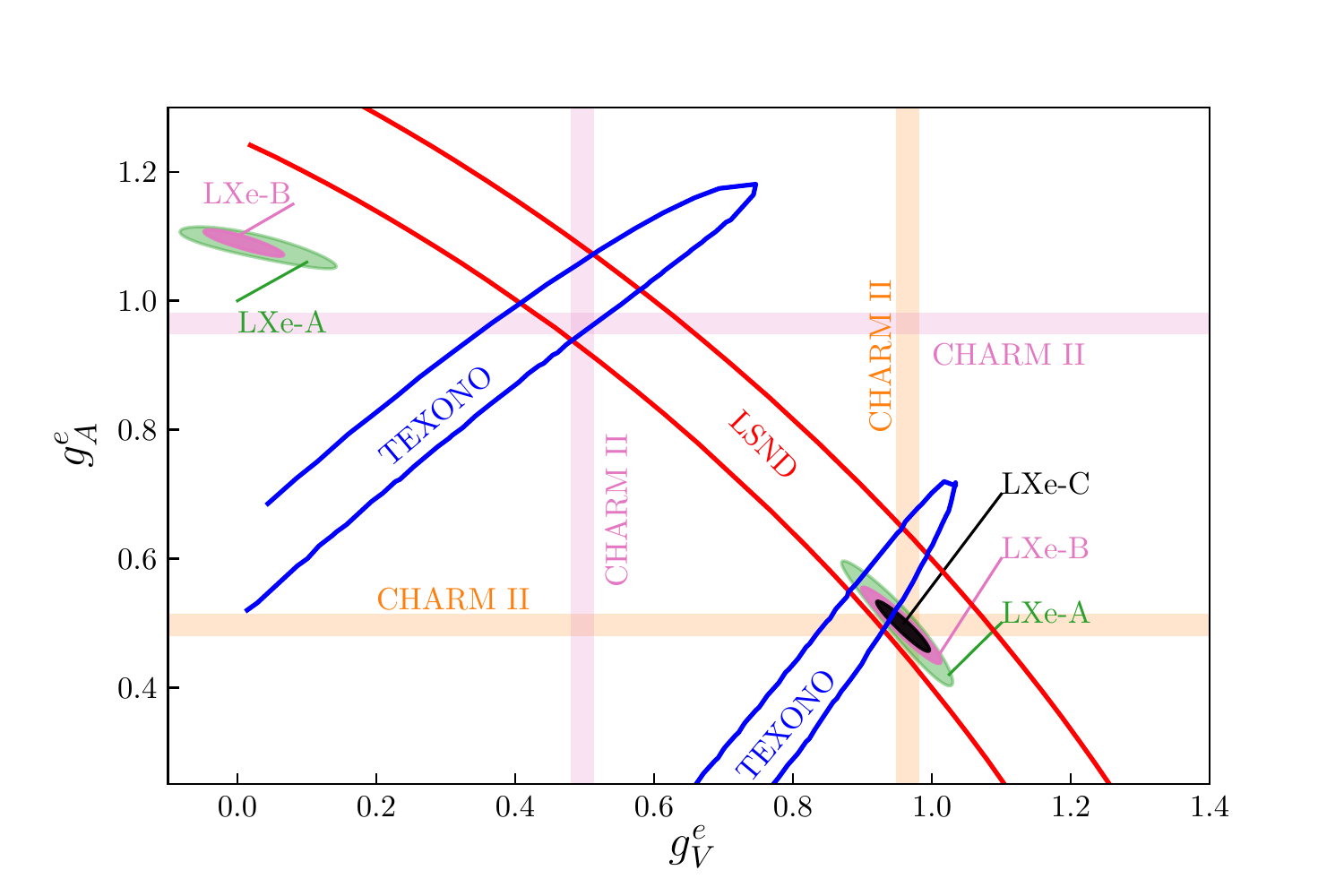}
\caption{\label{fig:gVgA}Sensitivity of $^{51}$Cr-LXe experiment to 
weak couplings parameters, $g_{V}^e$ and $g_{A}^e$, compared to existing constraints. 
The solid blobs represent the expected constraints, centered on the SM values, for the LXe-A, 
LXe-B, and LXe-B configurations in this work.  Prior constraints from
TEXONO (blue) and LSND (red) are taken from Fig.~10.1 of PDG \cite{Tanabashi:2018oca},
and the CHARM II constraint comes from Ref.~\cite{Vilain:1994qy}. }
\end{figure}

The $^{51}$Cr-LXe experiment would also provide a high precision
measurement of the weak coupling parameters $g_{V}^e$ and $g_{A}^e$.  We performed
a simultaneous $\chi^{2}$-fit of $g_{V}^e$ and $g_{A}^e$  based on Eqs.~(\ref{eq:LXe-7}) 
and (\ref{eq:LXe-8}). Fig.~\ref{fig:gVgA} shows the fit result in the $(g_{V}^e,\ g_{A}^e)$ 
plane, and compares it with the measurements from TEXONO, LSND, and CHARM II, where for 
$\boss{\nu}{\mu}$-$e$ scattering measurements of CHARM II, the weak 
coupling constants are converted using $g^{e}_{V}=g_V^{\mu}+1$ and $g^e_{A}=g_A^{\mu}+1$.  
Due to low statistics, TEXONO and LSND appear as long bands, which are mainly 
determined by the total event numbers with little impact from spectral information. 
In other words, TEXONO and LSND effectively measured only the total cross section 
since their low-statistics measurements were not sensitive to spectral distortions
from variations of $g_{V}$ and $g_{A}$. The CHARM II measurement, which has
much higher statistics, gives $g_{V}^{\mu}=-0.035\pm0.017$ and $g_{A}^{\mu}=-0.503\pm0.017$
(see the overlap of two orange bands). However, since the neutrino
energy is much higher than the electron mass in CHARM II, it also
has some discrete parameter degeneracy---the cross section is approximately
invariant when $(g_{V}^{\mu},\ g_{A}^{\mu})\rightarrow(g_{A}^{\mu},\ g_{V}^{\mu})$ and exactly
invariant under $(g_{V}^{\mu},\ g_{A}^{\mu})\rightarrow(-g_{V}^{\mu},\ -g_{A}^{\mu})$. 
For the $^{51}$Cr-LXe experiment, depending on the signal/background
ratio, similar situation may appear as well. This can be understood
by reformatting the cross section in powers of $T/E_{\nu}$:
\begin{equation}
\frac{d\sigma}{dT}=\frac{m_{e}G_{F}^{2}}{2\pi}\left[\left(g_{1}^{2}+g_{2}^{2}\right)-\left(2g_{2}^2+g_{1}g_{2}\frac{m_{e}}{E_{\nu}}\right)\frac{T}{E_{\nu}}+g_{2}^{2}\left(\frac{T}{E_{\nu}}\right)^{2}\right],\label{eq:LXe-31}
\end{equation}
which implies that for low recoil energies ($T/E_{\nu}\!\ll\!1$) we essentially measure only
$g_{1}^{2}+g_{2}^{2}$, while higher energy recoils are also sensitive to  
$2g_{2}^{2}+g_{1}g_{2}\frac{m_{e}}{E_{\nu}}$.  With these
two factors determined, $(g_{1},\ g_{2})$ would be approximately known
up to some discrete ambiguities. This explains why configurations
A and B have two separate low-$\Delta\chi^{2}$ regions, while for 
configuration C, this degeneracy is broken by the higher signal/background measurement, which is 
also sensitive to the quadratic term in $T/E_{\nu}$.  

\section{BSM neutrino interactions\label{sec:BSM}}

In many BSM theories, neutrinos have new interactions, which can be
categorized into two types, (i) interactions mediated by heavy particles,
and (ii) interactions mediated by light particles. 

For heavy mediators, one can integrate them out and study the effective
four-fermion interactions. As a model-independent approach, one can
write down all possible effective four-fermion interactions of neutrinos
with electrons (as we are considering elastic neutrino-electron scattering).
The most extensively studied effective interactions are the Non-Standard
Interactions (NSIs, reviewed in \cite{Davidson:2003ha,Ohlsson:2012kf,Farzan:2017xzy,Esteban:2018ppq})
which lead to interesting phenomenology in neutrino oscillations. In
addition, one can consider the most general four-fermion operators
with all possible Lorentz invariant forms considered, including Scalar,
Pseudoscalar, Vector, Axialvector, and Tensor (SPVAT) interactions
\cite{Healey:2013vka,Sevda:2016otj,Lindner:2016wff,Heurtier:2016otg,Rodejohann:2017vup,Kosmas:2017tsq,Magill:2017mps,Farzan:2018gtr,Yang:2018yvk,AristizabalSierra:2018eqm,Bischer:2018zcz,Brdar:2018qqj,Blaut:2018fis}.
We will study the sensitivity of a $^{51}$Cr-LXe
experiment to both frameworks.

For light mediators, the interactions can be much more multifarious.
As a demonstration of light mediator sensitivity, we will study a light $Z'$ 
with a very weak gauge coupling. Such a scenario is particular important
for dark matter studies~\cite{Feldman:2007wj,Dudas:2009uq,Buckley:2011vc,An:2012va,Arcadi:2013qia,Alves:2013tqa,Cline:2014dwa,Ducu:2015fda,Buchmueller:2014yoa,Alves:2015mua,Okada:2016gsh,Ge:2017mcq}
since additional $U(1)$ symmetries have been widely used to stabilize the
dark mater candidates.

\subsection{Non-Standard Interactions}

NSIs in $\nu$-$e$ scatting are usually formulated by the following Lagrangian:
\begin{equation}
{\cal L}_{{\rm NSI}}\supset2\sqrt{2}G_{F}\overline{\nu}_{\alpha}\gamma^{\mu}P_{L}\nu_{\beta}\overline{\psi_{e}}\gamma^{\mu}\left[\epsilon_{\alpha\beta}^{L}P_{L}+\epsilon_{\alpha\beta}^{R}P_{R}\right]\psi_{e},\label{eq:LXe-12}
\end{equation}
where $\psi_{e}$ is the Dirac spinor of the electron, $\epsilon_{\alpha\beta}^{L}$ and $\epsilon_{\alpha\beta}^{R}$
are flavor-dependent constants, and $P_{L/R}\equiv(1\mp\gamma^{5})/2$.
This Lagrangian originates in some models with heavy
vector bosons that interact with neutrinos and electrons (see, e.g.,
\cite{Heeck:2018nzc}). In addition, purely scalar interactions can
also generate NSIs at the loop level, and these loop-induced
NSI could potentially be very large in the presence of some secret
neutrino-scalar interactions~\cite{Bischer:2018zbd}. 

\begin{figure}
\centering

\includegraphics[width=7.5cm]{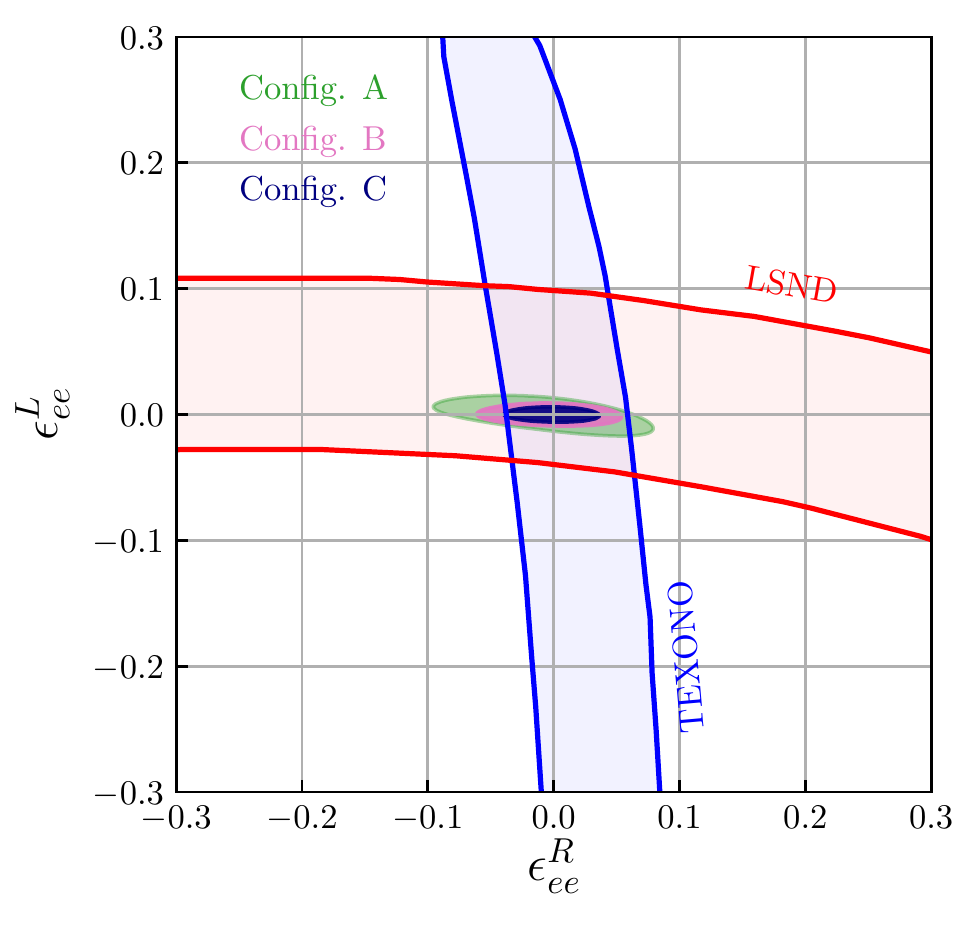}\includegraphics[width=7.5cm]{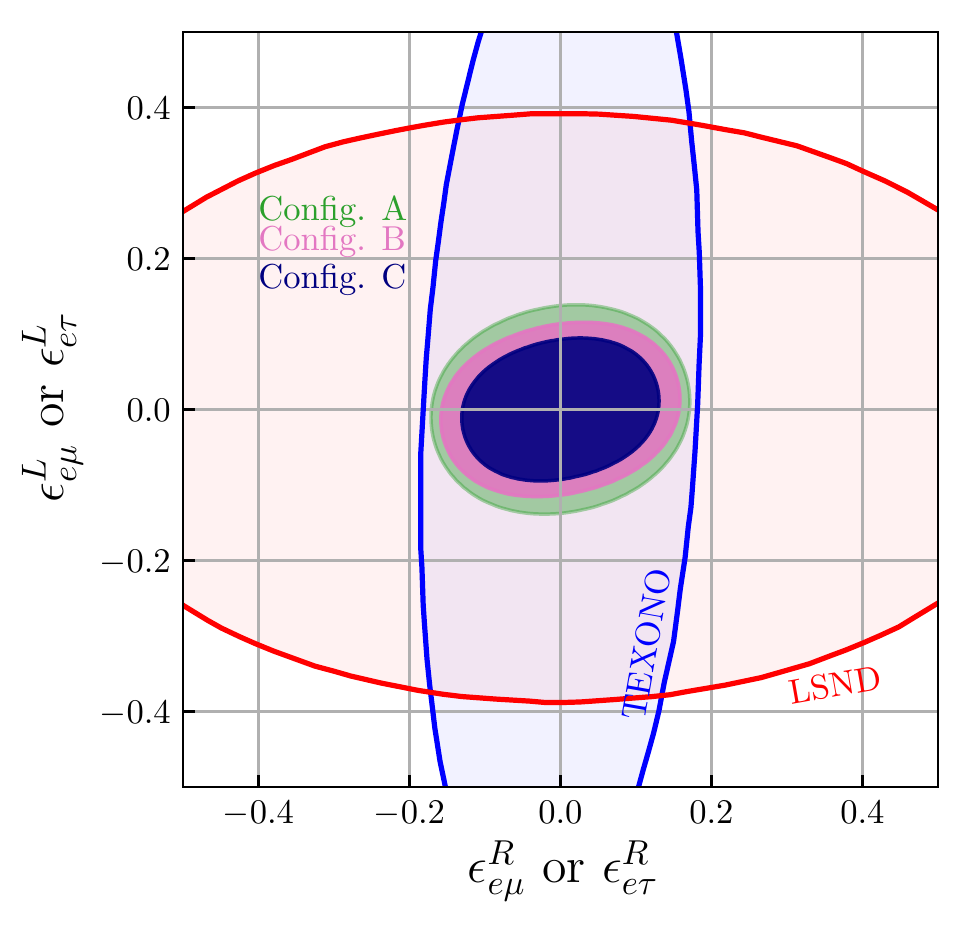}

\caption{\label{fig:NSI}Sensitivity at 90\% CL of a $^{51}$Cr-LXe experiment on the NSI parameters, compared to the existing bounds from the LSND and TEXONO experiments~\cite{Deniz:2010mp}. }
\end{figure}
Including NSIs, the cross section of $\nu_{e}$-$e$ scattering can
be derived by summing the cross sections of the three processes
$\nu_{e}+e\rightarrow\nu_{\alpha}+e$, where $\alpha=e$, $\mu$ or $\tau$,
which is
\begin{equation}
\frac{d\sigma}{dT}=\frac{m_{e}G_{F}^{2}}{2\pi}\left[\tilde{g}_{1}^{2}+\tilde{g}_{2}^{2}\left(1-\frac{T}{E_{\nu}}\right)^{2}-x_{12}\frac{m_{e}T}{E_{\nu}^{2}}\right],\label{eq:LXe-7-1}
\end{equation}
with $\tilde{g}_{1}^{2}$, $\tilde{g}_{1}^{2}$ and $x_{12}$ defined
as
\begin{align}
\tilde{g}_{1}^{2} & \equiv\left(g_{V}+g_{A}+2\epsilon_{ee}^{L}\right)^{2}+\sum_{\beta=\mu,\tau}\left(2\epsilon_{e\beta}^{L}\right)^{2},\label{eq:LXe-18}\\
\tilde{g}_{2}^{2} & \equiv\left(g_{V}-g_{A}+2\epsilon_{ee}^{R}\right)^{2}+\sum_{\beta=\mu,\tau}\left(2\epsilon_{e\beta}^{R}\right)^{2},\\
x_{12} & \equiv\left(g_{V}+g_{A}+2\epsilon_{ee}^{L}\right)\left(g_{V}-g_{A}+2\epsilon_{ee}^{R}\right)\nonumber \\
 & +\sum_{\beta=\mu,\tau}\left(2\epsilon_{e\beta}^{L}\right)\left(2\epsilon_{e\beta}^{R}\right).
\end{align}
Note that there is no interference between the three processes because
they have different flavor neutrinos in the final state. But for $\alpha=e$,
there is interference between the NSI and the SM interactions, which
significantly enhances the sensitivity to $\epsilon_{ee}^{L}$ and
$\epsilon_{ee}^{R}$.

\begin{table*}
\caption{\label{tab:NSI}Current NSI bounds from Ref.~\cite{Farzan:2017xzy}
compared with future $^{51}$Cr-LXe bounds (Configs. A, B, and
C). }

\begin{ruledtabular}
\begin{tabular}{cccc}
 & $\epsilon{}_{ee}^{P}$ & $\epsilon{}_{e\mu}^{P}$ & $\epsilon{}_{e\tau}^{P}$\tabularnewline
\hline 
known bounds ($P=L$) & $[-0.021,\ 0.052]$ & $[-0.13,\ 0.13]$ & $[-0.33,\ 0.33]$\tabularnewline
known bounds ($P=R$) & $[-0.07,\ 0.08]$ & $[-0.13,\ 0.13]$ & $[-0.19,\ 0.19]$\tabularnewline
\hline 
Config. A ($P=L$) & $[-0.017,\ 0.015]$ & $[-0.14,\ 0.14]$ & $[-0.14,\ 0.14]$\tabularnewline
Config. A ($P=R$) & $[-0.10,\ 0.08]$ & $[-0.17,\ 0.17]$ & $[-0.17,\ 0.17]$\tabularnewline
\hline 
Config. B ($P=L$) & $[-0.009,\ 0.009]$ & $[-0.12,\ 0.12]$ & $[-0.12,\ 0.12]$\tabularnewline
Config. B ($P=R$) & $[-0.06,\ 0.05]$ & $[-0.16,\ 0.16]$ & $[-0.16,\ 0.16]$\tabularnewline
\hline 
Config. C ($P=L$) & $[-0.006,\ 0.006]$ & $[-0.09,\ 0.09]$ & $[-0.09,\ 0.09]$\tabularnewline
Config. C ($P=R$) & $[-0.040,\ 0.036]$ & $[-0.13,\ 0.13]$ & $[-0.13,\ 0.13]$\tabularnewline
\end{tabular}
\end{ruledtabular}

\end{table*}

In Fig.~\ref{fig:NSI}, we present the result of a $\chi^{2}$-fit
analysis on the NSI sensitivity. The $\chi^{2}$-fit is performed
each time for a pair of $(\epsilon_{e\alpha}^{L},\ \epsilon_{e\alpha}^{R})$
with the other $\varepsilon$'s set to zero. Because the cross section
is completely symmetric under the $\mu$-$\tau$ exchange, $\nu_{e}$-$e$
scattering should have exactly the same sensitivities to $(\epsilon_{e\mu}^{L},\ \epsilon_{e\mu}^{R})$
and to $(\epsilon_{e\tau}^{L},\ \epsilon_{e\tau}^{R})$. So they are
shown in the same plot in the right panel. As we can see, $(\epsilon_{ee}^{L},\ \epsilon_{ee}^{R})$
would be stringently constrained by the $^{51}$Cr-LXe experiment
while $(\epsilon_{e\mu}^{L},\ \epsilon_{e\mu}^{R})$ and to $(\epsilon_{e\tau}^{L},\ \epsilon_{e\tau}^{R})$
would be less constrained due to the lack of interference.

The $^{51}$Cr-LXe bounds are compared to 
the current known bounds on NSIs. In general, direct measurements
of neutrino-electron scattering and measurements of neutrino oscillation
are sensitive to the NSIs considered here. In particular, there have
been bounds from LSND and TEXONO that can be readily superposed on
our results, as shown in Fig.~\ref{fig:NSI}. As for other bounds,
we refer to Ref.~\cite{Farzan:2017xzy} for the recently updated
summary, which have been included in Tab.~\ref{tab:NSI}. We conclude
that for most NSIs parameters, the $^{51}$Cr-LXe sensitivity to NSIs 
would generally exceed the current known bounds.

\subsection{SPVAT interactions}

More generally one can adopt an effective field
theory (EFT) approach to study BSM neutrino interactions and write
down all the possible Lorentz invariant operators as follows:
\begin{equation}
{\cal L}\supset\frac{G_{F}}{\sqrt{2}}\sum_{a=S,P,V,A,T}\overline{\nu}\Gamma^{a}\nu\left[\overline{\psi_{e}}\Gamma^{a}(C_{a}+D_{a}i_{a}\gamma^{5})\psi_{e}\right],\label{eq:SPVAT}
\end{equation}
with 
\begin{equation}
\left(\Gamma^{S},\ \Gamma^{P},\ \Gamma^{V},\ \Gamma^{A},\ \Gamma^{T}\right)\equiv\left(\mathbf{1},\ i\gamma^{5},\ \gamma^{\mu},\ \gamma^{\mu}\gamma^{5},\ \sigma^{\mu\nu}\equiv\frac{i}{2}[\gamma^{\mu},\gamma^{\nu}]\right).\label{eq:LXe-21}
\end{equation}
Here $C_{a}$ and $D_{a}$ are real constants if $i_{a}=i$ for $a=S,\ P$ and $T$; 
and $i_{a}=1$ for $a=V$ and $A$.  It should be noted that this reduces to the SM 
Lagrangian when $(C_{V},\ D_{V},\ C_{A},\ D_{A})=(g_{V},\ -g_{A},\ g_{A},\ -g_{V})$
and all other $C_a$ and $D_a$ are zero.

Regarding theoretical motivations, the first four types  ($a=S,\ P,\ V$ and $A$)
could originate from integrating out some heavy scalar or vector mediators.
The tensor interactions could be generated by integrating out heavy charged 
scalar mediators following necessary the Fierz transformations~\cite{Xu:2019dxe}. 
For Dirac neutrinos, all the SPVAT interactions could exist with
10 free parameters ($C_{a}$, $D_{a}$) in Eq.~(\ref{eq:SPVAT}).
For Majorana neutrinos, there are further constraints on the SPVAT
interactions such that the allowed parameter space is actually smaller, containing
only 6 free parameters \cite{Rodejohann:2017vup}. Such a difference
could be used to distinguish between Dirac and Majorana neutrinos
\cite{Rosen:1982pj}. In this work, we will ignore the additional
constraints of Majorana neutrinos and simply take the full parameter
spaces containing 10 parameters into consideration.

\begin{figure}
\centering

\includegraphics[width=7.5cm]{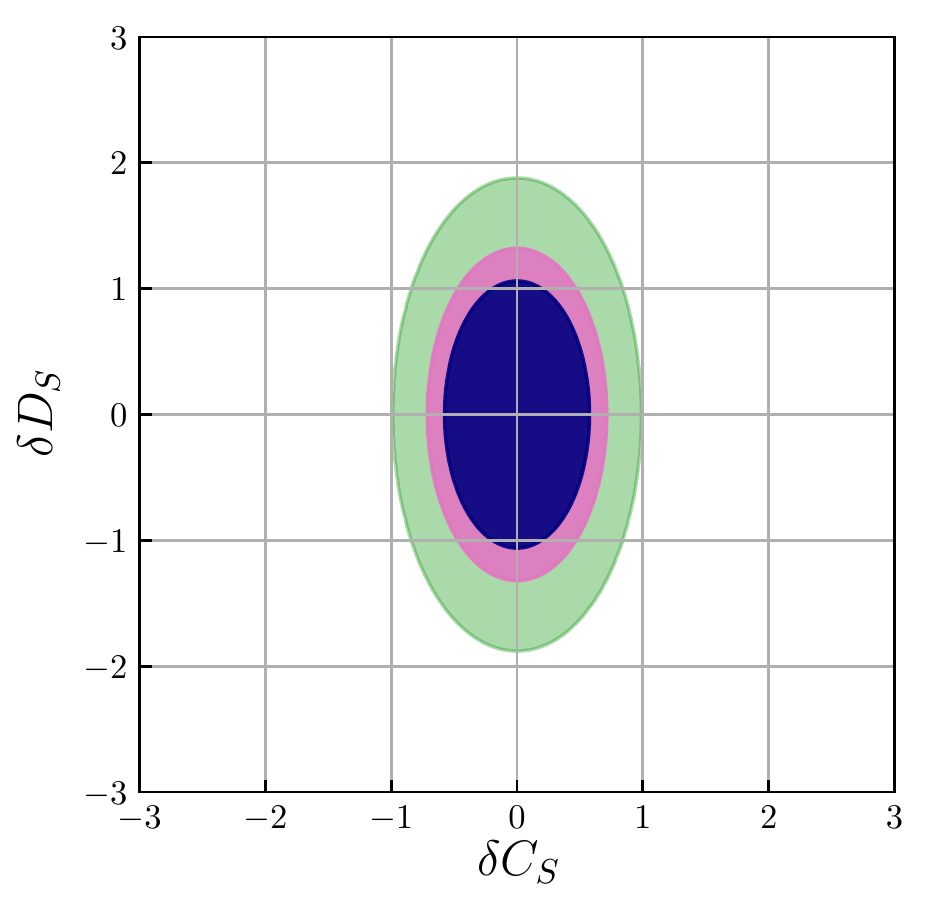}\includegraphics[width=7.5cm]{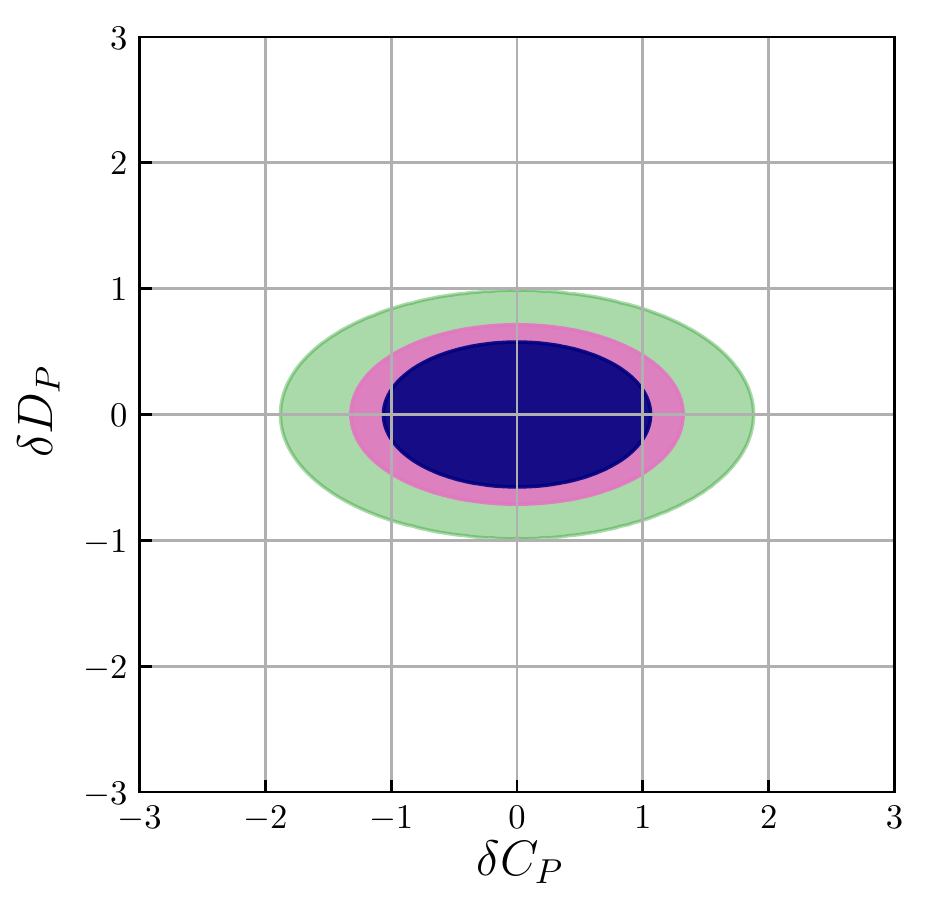}

\includegraphics[width=7.5cm]{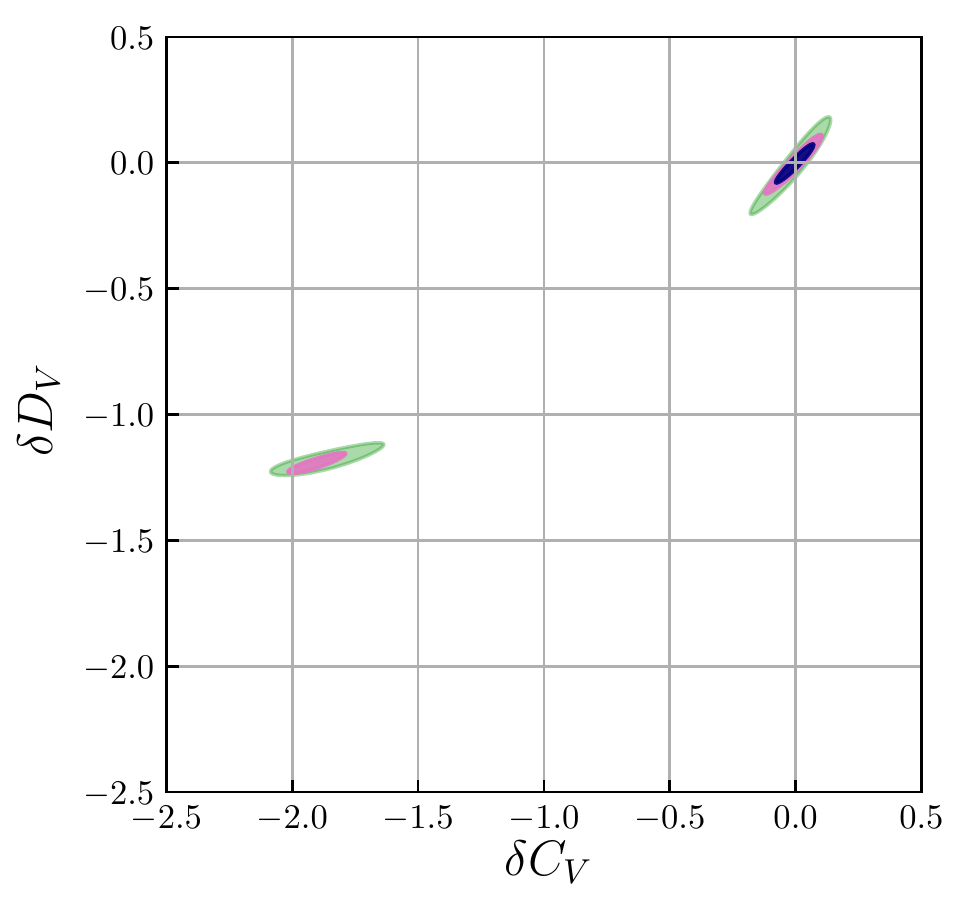}\includegraphics[width=7.5cm]{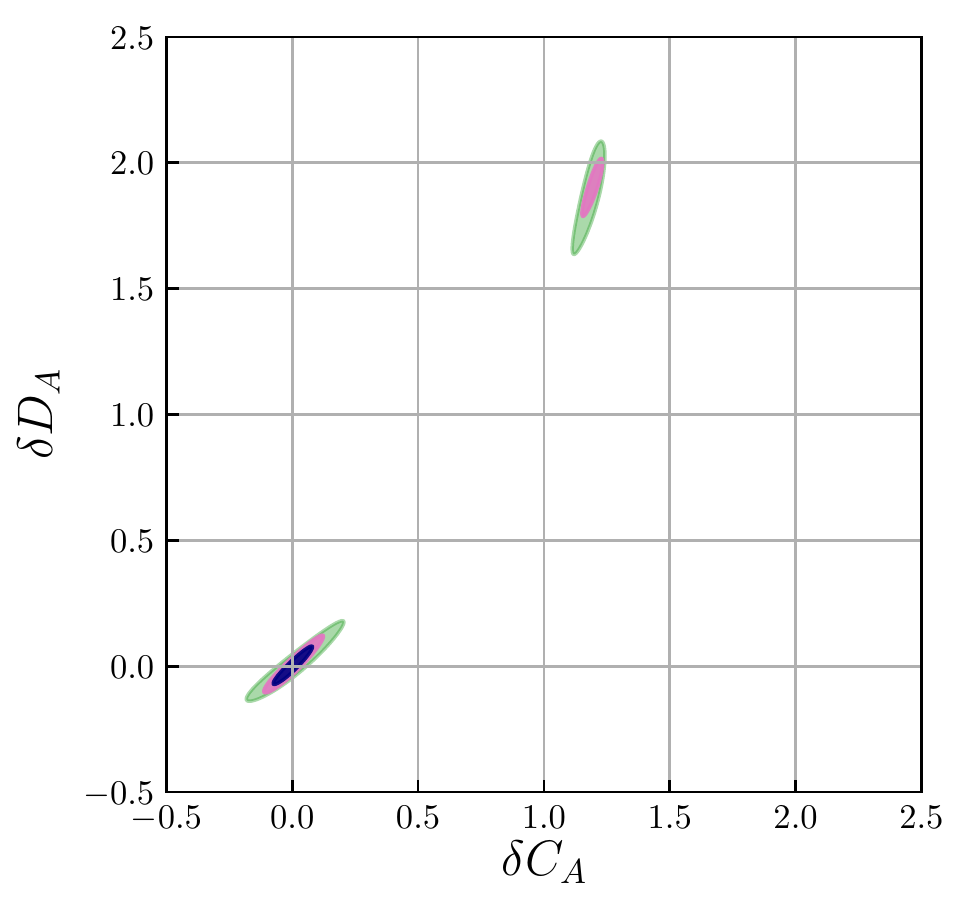}

\includegraphics[width=7.5cm]{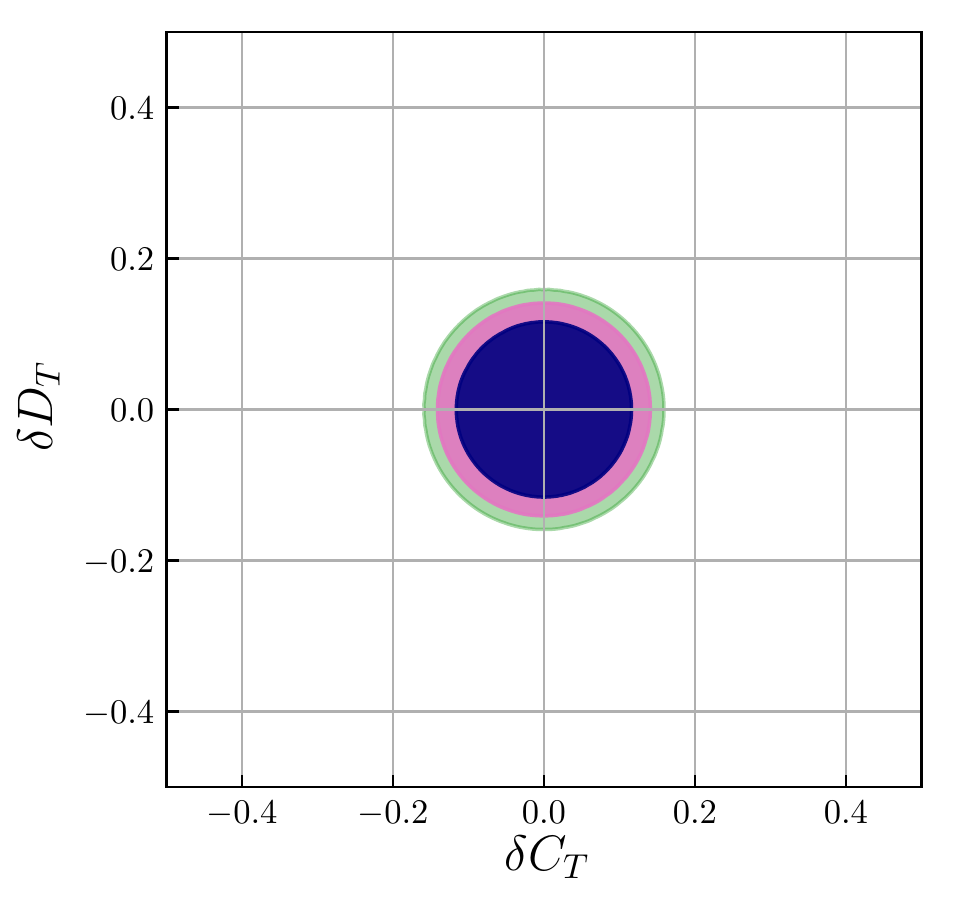}

\caption{\label{fig:SPVAT}Constraints on the SPVAT parameters. The color coding
is the same as Fig.~\ref{fig:NSI}.}
\end{figure}
In this model space, the cross section of $\nu_{e}$-$e$ scattering is given as follows
\cite{Rodejohann:2017vup}:
\begin{eqnarray}
\frac{d\sigma}{dT} & = & \frac{m_{e}G_{F}{}^{2}}{2\pi}\left[x_{1}+2x_{2}(1-\frac{T}{E_{\nu}})+x_{3}(1-\frac{T}{E_{\nu}})^{2}+x_{4}\frac{MT}{4E_{\nu}^{2}}\right],\label{eq:LXe-22}
\end{eqnarray}
where 
\begin{align}
x_{1} & \equiv\frac{1}{4}\left(C_{A}-D_{A}+C_{V}-D_{V}\right){}^{2}+\frac{1}{2}C_{P}C_{T}+\frac{1}{8}(C_{P}^{2}+C_{S}^{2}+D_{P}^{2}+D_{S}^{2})\nonumber \\
 & -\frac{1}{2}C_{S}C_{T}+C_{T}^{2}+\frac{1}{2}D_{P}D_{T}-\frac{1}{2}D_{S}D_{T}+D_{T}^{2},\label{eq:LXe-23}\\
x_{2} & \equiv-\frac{1}{8}\left(C_{P}^{2}+C_{S}^{2}+D_{P}^{2}+D_{S}^{2}\right)+C_{T}^{2}+D_{T}^{2},\label{eq:LXe-24}\\
x_{3} & \equiv\frac{1}{4}\left(C_{A}+D_{A}-C_{V}-D_{V}\right){}^{2}-\frac{1}{2}C_{P}C_{T}+\frac{1}{8}(C_{P}^{2}+C_{S}^{2}+D_{P}^{2}+D_{S}^{2})\nonumber \\
 & +\frac{1}{2}C_{T}C_{S}+C_{T}^{2}-\frac{1}{2}D_{P}D_{T}+\frac{1}{2}D_{S}D_{T}+D_{T}^{2},\label{eq:LXe-25}\\
x_{4} & \equiv-\left(C_{V}-D_{A}\right){}^{2}+\left(C_{A}-D_{V}\right){}^{2}+C_{S}^{2}-4C_{T}^{2}+D_{P}^{2}-4D_{T}^{2}.\label{eq:LXe-26}
\end{align}
In our analysis we are interested in extracting the non-SM part, so in our $\chi^2$-fit we define
\begin{equation}
(\delta C_{a},\ \delta D_{a})\equiv(C_{a},\ D_{a})-(C_{a},\ D_{a})^{{\rm SM}},\label{eq:LXe-32}
\end{equation}
and make a 2-parameter fit for each pair of $(\delta C_{a},\ \delta D_{a})$.
The result is presented in Fig.~\ref{fig:SPVAT}. For $a=S$ or $P$,
the $^{51}$Cr-LXe experiment would reach sensitivity around
$0.5\sim2$. For the other cases, the measurement could be even more precise.
 Note that for $a=V$ or $A$, there is parameter degeneracy which
would be removed in configuration C\@. The situation is similar to Fig.~\ref{fig:gVgA},
with the reason explained at the end of Sec.~\ref{sec:SM}.

\subsection{light $Z'$ }

\begin{figure}[t]
\centering

\includegraphics[width=10cm]{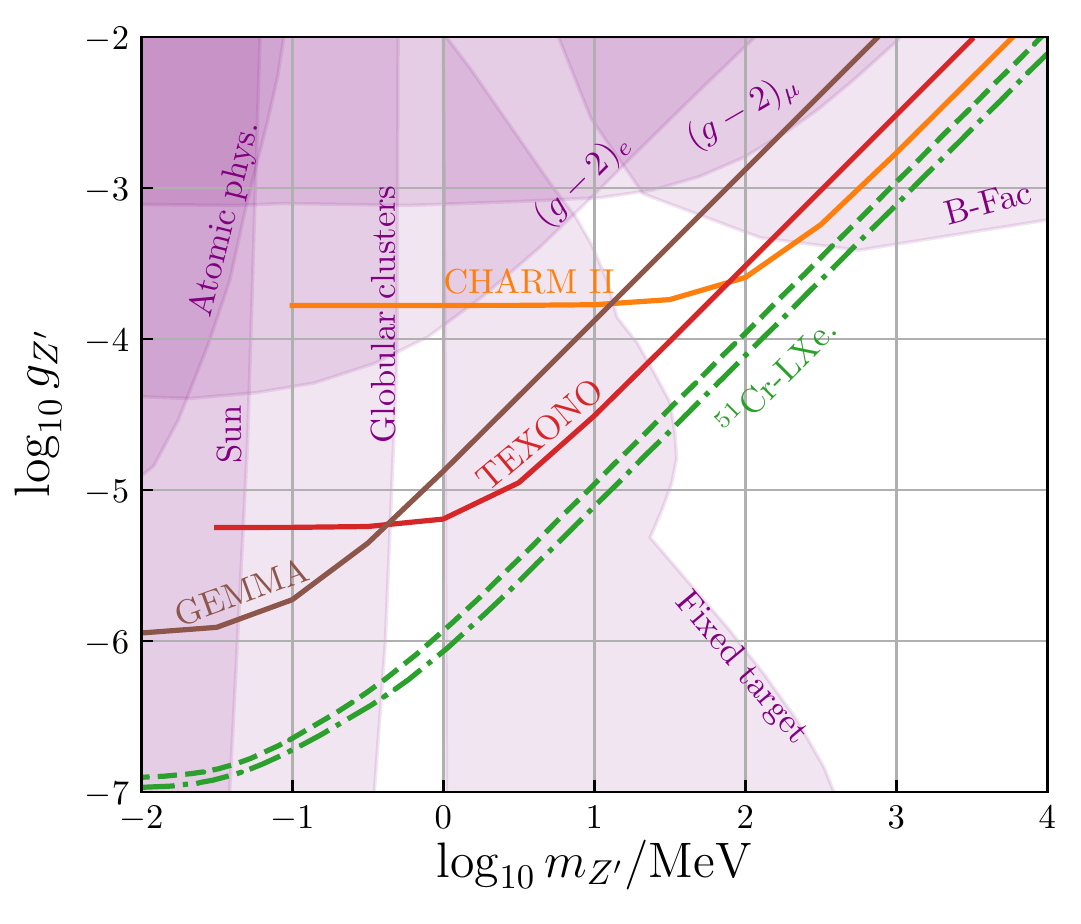}

\caption{\label{fig:light-Z} Constraints on the gauge coupling of light $Z'$. The constraints of
TEXONO, CHARM II, and GEMMA are taken from Ref.~\cite{Lindner:2018kjo},
other constraints are taken from Refs.~\cite{Harnik:2012ni,Bilmis:2015lja}.
The green dashed curve is for $^{51}$Cr-LXe with configuration
A, and dot-dashed for configuration C. Configuration B is not shown
here for simplicity, which should be between the dashed and dot-dashed
curves.}
\end{figure}

\begin{figure}[h]
\centering

\includegraphics[width=0.45\textwidth]{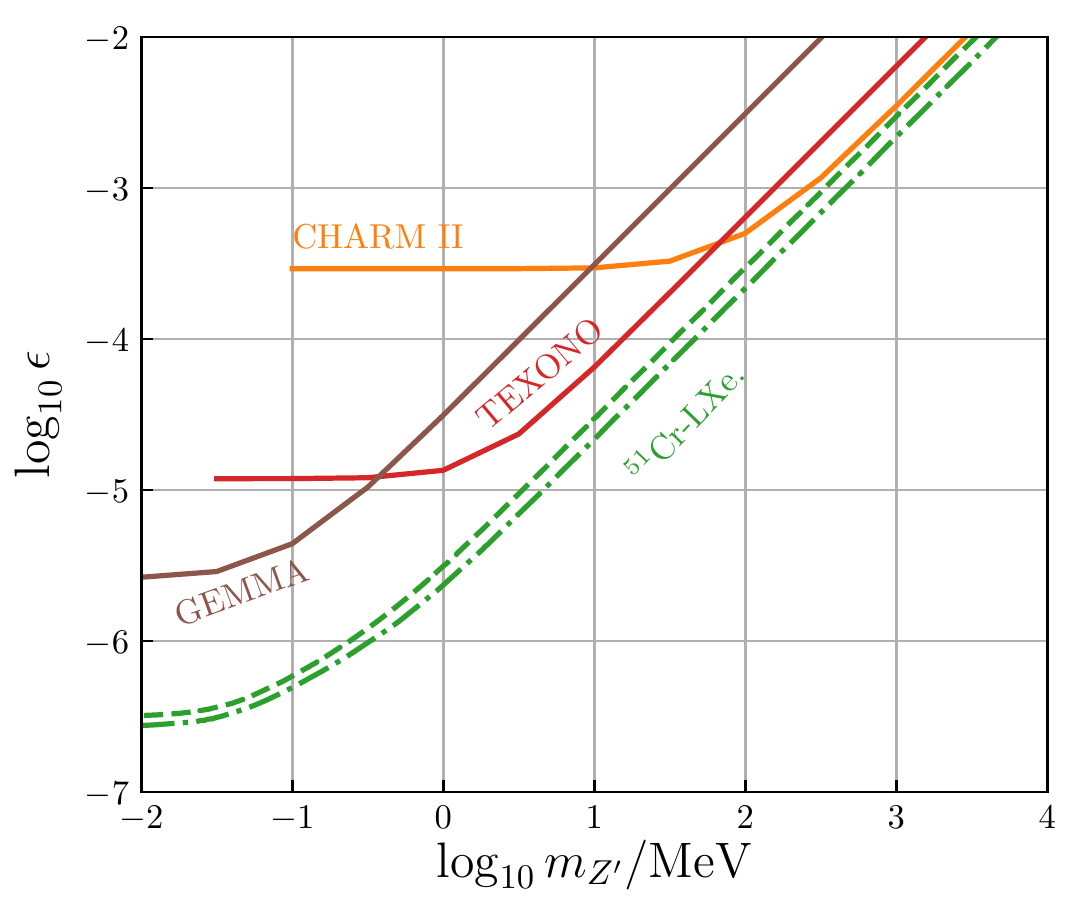}
\includegraphics[width=0.45\textwidth]{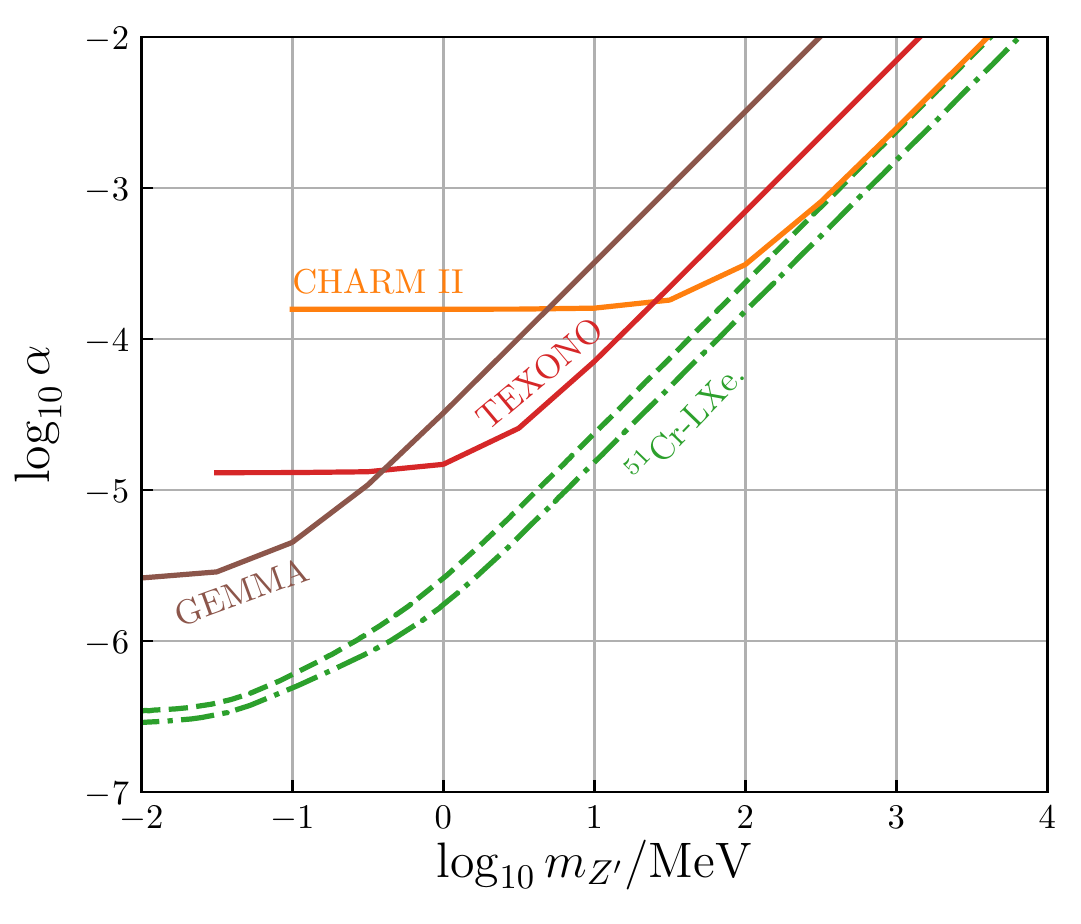}

\caption{\label{fig:light-Z-X} Constraints on the mass mixing $\alpha$ and kinetic mixing $\epsilon$ of light $Z'$. The constraints of
TEXONO, CHARM II, and GEMMA are taken from Ref.~\cite{Lindner:2018kjo}.
The green dashed curve is for $^{51}$Cr-LXe with configuration
A, and dot-dashed for configuration C. Configuration B is not shown
here for simplicity, which should be between the dashed and dot-dashed
curves.}
\end{figure}
Recently a generic light neutral vector boson (commonly referred to
as light $Z'$ ) has received extensive research interest due to its
connection with dark matter \cite{Feldman:2007wj,Dudas:2009uq,Buckley:2011vc,An:2012va,Arcadi:2013qia,Alves:2013tqa,Cline:2014dwa,Ducu:2015fda,Buchmueller:2014yoa,Alves:2015mua,Okada:2016gsh,Ge:2017mcq}.
Elastic neutrino scattering is intrinsically sensitive to light mediators
which usually show effect only in soft (low-momentum transfer) scattering
processes, where, for most other particles, the electromagnetic force
would be an overwhelming background. Due to their very weak SM interaction, 
neutrino scattering makes a very clean environment for probing new forces 
mediated by other light mediators~\cite{Bilmis:2015lja,Dutta:2015vwa,Dent:2016wcr,Cui:2017ytb,Farzan:2018gtr,Brdar:2018qqj,Bauer:2018onh,Abdullah:2018ykz,Lindner:2018kjo}.

The Lagrangian of light $Z'$ that concerns $\nu$-$e$ scattering
is generally formulated as follows 
\begin{equation}
{\cal L}\supset-\frac{g_{Z'}}{2}Z'_{\mu}\left[\overline{\nu}\gamma^{\mu}\nu+\overline{\ell}\gamma^{\mu}\ell\right]-\frac{1}{2}m_{Z'}^{2}Z'_{\mu}Z'^{\mu},\label{eq:LXe-29}
\end{equation}
where $g_{Z'}$ is the gauge coupling and $\ell$ is a charged lepton.
Depending on the models, there could be different charge assignments.
Here we simply assume the charge assignments
are the same as the $U(1)_{B-L}$ model (i.e., all leptons have the same
charge).
In generic $Z'$ models, the $Z'$ boson has both kinetic mixing 
and  mass mixing with the SM $Z^{0}$ boson.
As it has been studied in Ref.~\cite{Lindner:2016wff}, in the presence of both mixing, the $Z'$ and $Z^0$  interactions can be formulated as
\begin{equation}
{\cal L}\supset J_{Z}^{\mu}Z_{\mu}+J_{Z'}^{\mu}Z'_{\mu},\label{eq:Zp}
\end{equation}
where $ J_{Z}^{\mu}$ and $J_{Z'}^{\mu}$ are mixtures of the SM neutral current $J_{\rm NC}$ and a new current $J_{X}$:
\begin{eqnarray}
J_{Z}^{\mu} & = & c_{\alpha}J_{\text{NC}}^{\mu}-s_{\alpha}J_{X}^{\mu},\label{eq:Zp-3}\\
J_{Z'}^{\mu} & = & s_{\alpha}J_{\text{NC}}^{\mu}+c_{\alpha}J_{X}^{\mu}.
\label{eq:Zp-4}
\end{eqnarray}
Here  $\alpha$ is the mass mixing angle with $(c_{\alpha},\ s_{\alpha})\equiv(\cos\alpha,\ \sin\alpha)$. 
The leptonic part of the SM neutral current $J_{{\rm NC}}^{\mu}$
is
\begin{equation}
J_{{\rm NC}}^{\mu}=\frac{g}{c_{W}}\left[\overline{\nu}_{L}\gamma^{\mu}\frac{1}{2}\overline{\nu}_{L}+\overline{e}_{L}\gamma^{\mu}\frac{2s_{W}^{2}-1}{2}\overline{e}_{L}+\overline{e}_{R}\gamma^{\mu}s_{W}^{2}\overline{e}_{R}+\cdots\right],
\end{equation}
while the explicit form of $J_{X}^{\mu}$ is given by \cite{Lindner:2016wff}:
\begin{eqnarray}
J_{X}^{\mu} & = & \frac{1}{2c_{W}\sqrt{1-\epsilon^{2}}}\left[\overline{\nu}\gamma^{\mu}\left(-c_{W}g_{Z'}\right)\overline{\nu}+\overline{\nu}_{L}\gamma^{\mu}\left(g\epsilon s_{W}\right)\overline{\nu}_{L}\right.\nonumber \\
 &  & \left.+\overline{e}\gamma^{\mu}\left(-c_{W}g_{Z'}+g\epsilon s_{W}\right)\overline{e}+\overline{e}_{R}\gamma^{\mu}\left(g\epsilon s_{W}\right)\overline{e}_{R}+\cdots\right].\label{eq:nue-61-1}
\end{eqnarray}
The kinetic mixing $\epsilon$ is defined as $L\supset -\frac{1}{2}\epsilon F^{\mu\nu}F'_{\mu\nu}$ where $F^{\mu\nu}$ and $F'_{\mu\nu}$ are the gauge field tensors of the SM $U(1)_Y$ and the new $U(1)$ respectively.

The contribution of $Z'$ to the $\nu_{e}$-$e$ cross section can
be included by adding an energy-dependent term to $(g_{1},\ g_{2})$
in Eq.~(\ref{eq:LXe-7}):
\begin{equation}
\left(g_{1},\ g_{2}\right)\rightarrow\left(g_{1},\ g_{2}\right)+\frac{\left(\delta_{1},\ \delta_{2}\right)}{4\sqrt{2}G_{F}\left(2m_{e}T+m_{Z'}^{2}\right)},\label{eq:Zp-8}
\end{equation}
where $\left(\delta_{1},\ \delta_{2}\right)$ depends on $g_{Z'}$, $\epsilon$ and $\alpha$. When studying the effect of one of these three parameters, we assume the other two are negligibly small. Under this assumption, the expressions of  $\left(\delta_{1},\ \delta_{2}\right)$ are given as follows:
\begin{equation}
\left(\delta_{1},\ \delta_{2}\right)=\left(g_{Z'}^{2},\ g_{Z'}^{2}\right),\ {\rm for}\ g_{Z'}\neq0,\ \epsilon=\alpha=0,\label{eq:Zp-5}
\end{equation}
\begin{equation}
\left(\delta_{1},\ \delta_{2}\right)=\left(\frac{g^{2}s_{W}^{2}\epsilon^{2}}{c_{W}^{2}\left(1-\epsilon^{2}\right)},\ \frac{2g^{2}s_{W}^{2}\epsilon^{2}}{c_{W}^{2}\left(1-\epsilon^{2}\right)}\right),\ {\rm for}\ \epsilon\neq0,\ g_{Z'}=\alpha=0,\label{eq:Zp-6}
\end{equation}
\begin{equation}
\left(\delta_{1},\ \delta_{2}\right)=\left(\frac{g^{2}s_{\alpha}^{2}\left(2s_{W}^{2}-1\right)}{c_{W}^{2}},\ \frac{2g^{2}s_{\alpha}^{2}s_{W}^{2}}{c_{W}^{2}}\right),\ {\rm for}\ \alpha\neq0,\ g_{Z'}=\epsilon=0.\label{eq:Zp-7}
\end{equation}



Using Eqs.~(\ref{eq:Zp-8}) to (\ref{eq:Zp-7}), we study the sensitivity of the Cr-LXe experiment to the three  parameters $g_{Z'}$,  $\epsilon$,  $\alpha$ and present the results in Figs.~\ref{fig:light-Z} and \ref{fig:light-Z-X}, in comparison with  other known bounds taken from Refs.~\cite{Harnik:2012ni,Bilmis:2015lja,Lindner:2018kjo}.
For  the gauge coupling $g_{Z'}$ shown in Fig.~\ref{fig:light-Z}, in general, neutrino-electron
scattering experiments (CHARM II, TEXONO, and GEMMA) provide the leading
constraints prior to some constraints obtained from
atomic physics and measurements of muon and electron anomalous
magnetic moments, $(g-2)_{e}$ and $(g-2)_{\mu}$. They are also complementary
to some other constraints from fixed target experiments, astrophysical
observations (energy loss in the sun and globular clusters), and the B-Factories.
Due to its low threshold and high statistics, the $^{51}$Cr-LXe
experiment would significantly improve upon existing constraints
from neutrino-electron scattering. In the low mass limit, the $^{51}$Cr-LXe
experiment has sensitivity to small $g_{Z'}$ down to $10^{-7}$.
For heavy $Z'$, it would  exceed the CHARM II experiment which is
based on high-energy neutrino beams. 
The bounds on the kinetic mixing $\epsilon$ and the mass mixing $\alpha$ in Fig.~\ref{fig:light-Z-X} are similar to the $g_{Z'}$ bounds. Even with configuration A,  the $^{51}$Cr-LXe
experiment could provide the most stringent constraints on all these $Z'$ parameters, showing its the great potential in constraining or probing the new physics of light $Z'$.

\section{Conclusion\label{sec:Conclusion}}

In this work, we have studied the potential for $\nu_{e}$-$e$ scattering in a LXe
dark matter detector using $^{51}{\rm Cr}$ as a radioactive neutrino
source ($^{51}$Cr-LXe). Assuming a 5~MCi $^{51}{\rm Cr}$ source
located 1~m below a 6-ton LXe detector with the current state-of-the-art background
and running for 100 days (configuration A), one can already achieve
a high statistics ($1.2\times10^{4}$ signal events) measurement of
$\nu_{e}$-$e$ scattering. However, only the low energy part ($T\lesssim100$ keV)
of the recoil spectrum can be efficiently measured while the higher
energy signal would be submerged by the background, as shown in Fig.~\ref{fig:Signal}. 
To enhance the signal/background ratio at high $T$ so that the full recoil spectrum
($0<T\lesssim559$ keV) can be measured, the background must
be reduced by, for example, using $^{136}{\rm Xe}$-depleted LXe and a shorter
exposure time. For this purpose, we also considered configurations B
and C (see Tab.~\ref{tab:t}) for which the signal is greater than the background over the whole
spectrum (see Fig.~\ref{fig:Signal}).

Based on these three configurations, we study low energy precision measurement
of the SM parameters $\sin^{2}\theta_{W}$, $g_{V}$ and $g_{A}$. For $\sin^{2}\theta_{W}$,
we find that $1\sim2\%$ precision can be attained, as presented in
Fig.~\ref{fig:sw}, and Tab.~\ref{tab:sw}. As for $g_{V}$ and $g_{A})$,
such an experiment would provide the most precise measurement among $\boss{\nu}{e}$
scattering experiments (see Fig.~\ref{fig:gVgA}), and be comparable to
the CHARM II measurements based on $\boss{\nu}{\mu}$ 
scattering.  Due to this high precision, we expect that the $^{51}$Cr-LXe
experiment would be excellent at constraining BSM neutrino interactions.
Indeed, according to our analyses, it would generate leading or complementary
constraints on NSIs, SPVAT interactions, and light $Z'$, as presented
in Figs.~\ref{fig:NSI}, \ref{fig:SPVAT}, and \ref{fig:light-Z}
respectively. For example, take NSIs, the most extensively studied scenario: the 
$^{51}$Cr-LXe experiment is sensitive to six of the NSI parameters, namely 
$\epsilon_{ee}^{L}$, $\epsilon_{ee}^{R}$, $\epsilon_{e\mu}^{L}$, $\epsilon_{e\mu}^{R}$, 
$\epsilon_{e\tau}^{L}$, and $\epsilon_{e\tau}^{R}$.  In Tab.~\ref{tab:NSI}
and Fig.~\ref{fig:NSI} we show that $\epsilon_{e\mu}^{L}$, $\epsilon_{e\mu}^{R}$, 
$\epsilon_{e\tau}^{L}$, and $\epsilon_{e\tau}^{R}$ would be constrained to the 
limit around $0.1\sim0.2$, and $\epsilon_{ee}^{L}$ and $\epsilon_{ee}^{R}$
can be more stringently constrained due to the interference with the
SM signal, sometimes even down to $6\times10^{-3}$. These constraints
would generally exceed the existing bounds, which would be important
for future long baseline experiments such as DUNE and T2HK\@. In addition,
probing SPVAT interactions and light $Z'$ in $\nu_{e}$-$e$ scattering
would shed light on the fundamental theories of both neutrinos and
dark matter.

\begin{acknowledgments}
We thank Werner Rodejohann for helpful discussions.  This work was supported by the Max Planck Gesellschaft (MPG),
and the U.S. Department of Energy under grant number DE-SC0018327.
\end{acknowledgments}

\appendix

\section{Geometrical calculation of $r_{{\rm avg}}$\label{sec:Geometrical}}

According to the definition of $r_{{\rm avg}}$ in Eq.~(\ref{eq:r_eff}),
we have
\begin{align}
\frac{1}{r_{{\rm avg}}^{2}}=\frac{1}{V}\int dV\frac{1}{r^{2}} & =\frac{1}{V}\int\frac{1}{r^{2}}dxdydz,\label{eq:LXe-13}
\end{align}
where the volume is
\begin{equation}
V=\pi(d/2)^{2}h.\label{eq:LXe-15}
\end{equation}
Here the Cartesian coordinate $x$-$y$-$z$ is set in such a way
that $z$ is the cylindrical axis the $x$-$y$ is the horizontal
plane. By defining
\begin{equation}
R^{2}\equiv x^{2}+y^{2},\label{eq:LXe-14}
\end{equation}
we can compute the integral in the cylindrical coordinate system:
\begin{align}
\int\frac{1}{r^{2}}dxdydz & =\int\frac{R}{(R\cos\theta)^{2}+(R\sin\theta)^{2}+z{}^{2}}dzdRd\theta\nonumber \\
 & =\int_{0}^{d/2}dR\int_{L}^{L+h}\frac{2\pi R}{R^{2}+z{}^{2}}dz\nonumber \\
 & =\pi d\tan^{-1}\left(\frac{2(h+L)}{d}\right)-\pi d\tan^{-1}\left(\frac{2L}{d}\right)\nonumber \\
 & +\pi L\log\frac{L^{2}\left(d^{2}+4(h+L)^{2}\right)}{\left(d^{2}+4L^{2}\right)(h+L)^{2}}+\pi h\log\frac{d^{2}+4(h+L)^{2}}{4(h+L)^{2}}.\label{eq:int_r}
\end{align}
Therefore, the result is
\begin{equation}
r_{{\rm avg}}=\left[\frac{4L}{d^{2}h}\log\frac{L^{2}\left(d^{2}+4(h+L)^{2}\right)}{\left(d^{2}+4L^{2}\right)(h+L)^{2}}+\frac{4}{d^{2}}\log\frac{d^{2}+4(h+L)^{2}}{4(h+L)^{2}}-\frac{4}{dh}\tan^{-1}\left(\frac{2L}{d}\right)+\frac{4}{dh}\tan^{-1}\left(\frac{2(h+L)}{d}\right)\right]^{-\frac{1}{2}}.\label{eq:r_eff_analytic}
\end{equation}

\bibliographystyle{apsrev4-1}
\bibliography{ref}

\begin{thebibliography}{94}%
\makeatletter
\providecommand \@ifxundefined [1]{%
 \@ifx{#1\undefined}
}%
\providecommand \@ifnum [1]{%
 \ifnum #1\expandafter \@firstoftwo
 \else \expandafter \@secondoftwo
 \fi
}%
\providecommand \@ifx [1]{%
 \ifx #1\expandafter \@firstoftwo
 \else \expandafter \@secondoftwo
 \fi
}%
\providecommand \natexlab [1]{#1}%
\providecommand \enquote  [1]{``#1''}%
\providecommand \bibnamefont  [1]{#1}%
\providecommand \bibfnamefont [1]{#1}%
\providecommand \citenamefont [1]{#1}%
\providecommand \href@noop [0]{\@secondoftwo}%
\providecommand \href [0]{\begingroup \@sanitize@url \@href}%
\providecommand \@href[1]{\@@startlink{#1}\@@href}%
\providecommand \@@href[1]{\endgroup#1\@@endlink}%
\providecommand \@sanitize@url [0]{\catcode `\\12\catcode `\$12\catcode
  `\&12\catcode `\#12\catcode `\^12\catcode `\_12\catcode `\%12\relax}%
\providecommand \@@startlink[1]{}%
\providecommand \@@endlink[0]{}%
\providecommand \url  [0]{\begingroup\@sanitize@url \@url }%
\providecommand \@url [1]{\endgroup\@href {#1}{\urlprefix }}%
\providecommand \urlprefix  [0]{URL }%
\providecommand \Eprint [0]{\href }%
\providecommand \doibase [0]{http://dx.doi.org/}%
\providecommand \selectlanguage [0]{\@gobble}%
\providecommand \bibinfo  [0]{\@secondoftwo}%
\providecommand \bibfield  [0]{\@secondoftwo}%
\providecommand \translation [1]{[#1]}%
\providecommand \BibitemOpen [0]{}%
\providecommand \bibitemStop [0]{}%
\providecommand \bibitemNoStop [0]{.\EOS\space}%
\providecommand \EOS [0]{\spacefactor3000\relax}%
\providecommand \BibitemShut  [1]{\csname bibitem#1\endcsname}%
\let\auto@bib@innerbib\@empty
\bibitem [{\citenamefont {Davidson}\ \emph {et~al.}(2003)\citenamefont
  {Davidson}, \citenamefont {Pena-Garay}, \citenamefont {Rius},\ and\
  \citenamefont {Santamaria}}]{Davidson:2003ha}%
  \BibitemOpen
  \bibfield  {author} {\bibinfo {author} {\bibfnamefont {S.}~\bibnamefont
  {Davidson}}, \bibinfo {author} {\bibfnamefont {C.}~\bibnamefont
  {Pena-Garay}}, \bibinfo {author} {\bibfnamefont {N.}~\bibnamefont {Rius}}, \
  and\ \bibinfo {author} {\bibfnamefont {A.}~\bibnamefont {Santamaria}},\
  }\href {\doibase 10.1088/1126-6708/2003/03/011} {\bibfield  {journal}
  {\bibinfo  {journal} {JHEP}\ }\textbf {\bibinfo {volume} {03}},\ \bibinfo
  {pages} {011} (\bibinfo {year} {2003})},\ \Eprint
  {http://arxiv.org/abs/hep-ph/0302093} {arXiv:hep-ph/0302093 [hep-ph]}
  \BibitemShut {NoStop}%
\bibitem [{\citenamefont {Ohlsson}(2013)}]{Ohlsson:2012kf}%
  \BibitemOpen
  \bibfield  {author} {\bibinfo {author} {\bibfnamefont {T.}~\bibnamefont
  {Ohlsson}},\ }\href {\doibase 10.1088/0034-4885/76/4/044201} {\bibfield
  {journal} {\bibinfo  {journal} {Rept. Prog. Phys.}\ }\textbf {\bibinfo
  {volume} {76}},\ \bibinfo {pages} {044201} (\bibinfo {year} {2013})},\
  \Eprint {http://arxiv.org/abs/1209.2710} {arXiv:1209.2710 [hep-ph]}
  \BibitemShut {NoStop}%
\bibitem [{\citenamefont {Farzan}\ and\ \citenamefont
  {Tortola}(2018)}]{Farzan:2017xzy}%
  \BibitemOpen
  \bibfield  {author} {\bibinfo {author} {\bibfnamefont {Y.}~\bibnamefont
  {Farzan}}\ and\ \bibinfo {author} {\bibfnamefont {M.}~\bibnamefont
  {Tortola}},\ }\href {\doibase 10.3389/fphy.2018.00010} {\bibfield  {journal}
  {\bibinfo  {journal} {Front.in Phys.}\ }\textbf {\bibinfo {volume} {6}},\
  \bibinfo {pages} {10} (\bibinfo {year} {2018})},\ \Eprint
  {http://arxiv.org/abs/1710.09360} {arXiv:1710.09360 [hep-ph]} \BibitemShut
  {NoStop}%
\bibitem [{\citenamefont {Esteban}\ \emph {et~al.}(2018)\citenamefont
  {Esteban}, \citenamefont {Gonzalez-Garcia}, \citenamefont {Maltoni},
  \citenamefont {Martinez-Soler},\ and\ \citenamefont
  {Salvado}}]{Esteban:2018ppq}%
  \BibitemOpen
  \bibfield  {author} {\bibinfo {author} {\bibfnamefont {I.}~\bibnamefont
  {Esteban}}, \bibinfo {author} {\bibfnamefont {M.~C.}\ \bibnamefont
  {Gonzalez-Garcia}}, \bibinfo {author} {\bibfnamefont {M.}~\bibnamefont
  {Maltoni}}, \bibinfo {author} {\bibfnamefont {I.}~\bibnamefont
  {Martinez-Soler}}, \ and\ \bibinfo {author} {\bibfnamefont {J.}~\bibnamefont
  {Salvado}},\ }\href@noop {} {\  (\bibinfo {year} {2018})},\ \Eprint
  {http://arxiv.org/abs/1805.04530} {arXiv:1805.04530 [hep-ph]} \BibitemShut
  {NoStop}%
\bibitem [{\citenamefont {Forero}\ and\ \citenamefont
  {Huber}(2016)}]{Forero:2016cmb}%
  \BibitemOpen
  \bibfield  {author} {\bibinfo {author} {\bibfnamefont {D.~V.}\ \bibnamefont
  {Forero}}\ and\ \bibinfo {author} {\bibfnamefont {P.}~\bibnamefont {Huber}},\
  }\href {\doibase 10.1103/PhysRevLett.117.031801} {\bibfield  {journal}
  {\bibinfo  {journal} {Phys. Rev. Lett.}\ }\textbf {\bibinfo {volume} {117}},\
  \bibinfo {pages} {031801} (\bibinfo {year} {2016})},\ \Eprint
  {http://arxiv.org/abs/1601.03736} {arXiv:1601.03736 [hep-ph]} \BibitemShut
  {NoStop}%
\bibitem [{\citenamefont {Miranda}\ \emph {et~al.}(2016)\citenamefont
  {Miranda}, \citenamefont {Tortola},\ and\ \citenamefont
  {Valle}}]{Miranda:2016wdr}%
  \BibitemOpen
  \bibfield  {author} {\bibinfo {author} {\bibfnamefont {O.~G.}\ \bibnamefont
  {Miranda}}, \bibinfo {author} {\bibfnamefont {M.}~\bibnamefont {Tortola}}, \
  and\ \bibinfo {author} {\bibfnamefont {J.~W.~F.}\ \bibnamefont {Valle}},\
  }\href {\doibase 10.1103/PhysRevLett.117.061804} {\bibfield  {journal}
  {\bibinfo  {journal} {Phys. Rev. Lett.}\ }\textbf {\bibinfo {volume} {117}},\
  \bibinfo {pages} {061804} (\bibinfo {year} {2016})},\ \Eprint
  {http://arxiv.org/abs/1604.05690} {arXiv:1604.05690 [hep-ph]} \BibitemShut
  {NoStop}%
\bibitem [{\citenamefont {Masud}\ \emph {et~al.}(2016)\citenamefont {Masud},
  \citenamefont {Chatterjee},\ and\ \citenamefont {Mehta}}]{Masud:2015xva}%
  \BibitemOpen
  \bibfield  {author} {\bibinfo {author} {\bibfnamefont {M.}~\bibnamefont
  {Masud}}, \bibinfo {author} {\bibfnamefont {A.}~\bibnamefont {Chatterjee}}, \
  and\ \bibinfo {author} {\bibfnamefont {P.}~\bibnamefont {Mehta}},\ }\href
  {\doibase 10.1088/0954-3899/43/9/095005/meta, 10.1088/0954-3899/43/9/095005}
  {\bibfield  {journal} {\bibinfo  {journal} {J. Phys.}\ }\textbf {\bibinfo
  {volume} {G43}},\ \bibinfo {pages} {095005} (\bibinfo {year} {2016})},\
  \Eprint {http://arxiv.org/abs/1510.08261} {arXiv:1510.08261 [hep-ph]}
  \BibitemShut {NoStop}%
\bibitem [{\citenamefont {Bakhti}\ and\ \citenamefont
  {Farzan}(2016)}]{Bakhti:2016prn}%
  \BibitemOpen
  \bibfield  {author} {\bibinfo {author} {\bibfnamefont {P.}~\bibnamefont
  {Bakhti}}\ and\ \bibinfo {author} {\bibfnamefont {Y.}~\bibnamefont
  {Farzan}},\ }\href {\doibase 10.1007/JHEP07(2016)109} {\bibfield  {journal}
  {\bibinfo  {journal} {JHEP}\ }\textbf {\bibinfo {volume} {07}},\ \bibinfo
  {pages} {109} (\bibinfo {year} {2016})},\ \Eprint
  {http://arxiv.org/abs/1602.07099} {arXiv:1602.07099 [hep-ph]} \BibitemShut
  {NoStop}%
\bibitem [{\citenamefont {Masud}\ and\ \citenamefont
  {Mehta}(2016)}]{Masud:2016bvp}%
  \BibitemOpen
  \bibfield  {author} {\bibinfo {author} {\bibfnamefont {M.}~\bibnamefont
  {Masud}}\ and\ \bibinfo {author} {\bibfnamefont {P.}~\bibnamefont {Mehta}},\
  }\href {\doibase 10.1103/PhysRevD.94.013014} {\bibfield  {journal} {\bibinfo
  {journal} {Phys. Rev.}\ }\textbf {\bibinfo {volume} {D94}},\ \bibinfo {pages}
  {013014} (\bibinfo {year} {2016})},\ \Eprint
  {http://arxiv.org/abs/1603.01380} {arXiv:1603.01380 [hep-ph]} \BibitemShut
  {NoStop}%
\bibitem [{\citenamefont {C}\ and\ \citenamefont {Mohanta}(2017)}]{C.:2017yqh}%
  \BibitemOpen
  \bibfield  {author} {\bibinfo {author} {\bibfnamefont {S.}~\bibnamefont {C}}\
  and\ \bibinfo {author} {\bibfnamefont {R.}~\bibnamefont {Mohanta}},\ }\href
  {\doibase 10.1140/epjc/s10052-017-4600-8} {\bibfield  {journal} {\bibinfo
  {journal} {Eur. Phys. J.}\ }\textbf {\bibinfo {volume} {C77}},\ \bibinfo
  {pages} {32} (\bibinfo {year} {2017})},\ \Eprint
  {http://arxiv.org/abs/1701.00327} {arXiv:1701.00327 [hep-ph]} \BibitemShut
  {NoStop}%
\bibitem [{\citenamefont {Hyde}(2018)}]{Hyde:2018tqt}%
  \BibitemOpen
  \bibfield  {author} {\bibinfo {author} {\bibfnamefont {J.~M.}\ \bibnamefont
  {Hyde}},\ }\href@noop {} {\  (\bibinfo {year} {2018})},\ \Eprint
  {http://arxiv.org/abs/1806.09221} {arXiv:1806.09221 [hep-ph]} \BibitemShut
  {NoStop}%
\bibitem [{\citenamefont {Deepthi}\ \emph {et~al.}(2018)\citenamefont
  {Deepthi}, \citenamefont {Goswami},\ and\ \citenamefont
  {Nath}}]{Deepthi:2017gxg}%
  \BibitemOpen
  \bibfield  {author} {\bibinfo {author} {\bibfnamefont {K.~N.}\ \bibnamefont
  {Deepthi}}, \bibinfo {author} {\bibfnamefont {S.}~\bibnamefont {Goswami}}, \
  and\ \bibinfo {author} {\bibfnamefont {N.}~\bibnamefont {Nath}},\ }\href
  {\doibase 10.1016/j.nuclphysb.2018.09.004} {\bibfield  {journal} {\bibinfo
  {journal} {Nucl. Phys.}\ }\textbf {\bibinfo {volume} {B936}},\ \bibinfo
  {pages} {91} (\bibinfo {year} {2018})},\ \Eprint
  {http://arxiv.org/abs/1711.04840} {arXiv:1711.04840 [hep-ph]} \BibitemShut
  {NoStop}%
\bibitem [{\citenamefont {Konetschny}\ and\ \citenamefont
  {Kummer}(1977)}]{Konetschny:1977bn}%
  \BibitemOpen
  \bibfield  {author} {\bibinfo {author} {\bibfnamefont {W.}~\bibnamefont
  {Konetschny}}\ and\ \bibinfo {author} {\bibfnamefont {W.}~\bibnamefont
  {Kummer}},\ }\href {\doibase 10.1016/0370-2693(77)90407-5} {\bibfield
  {journal} {\bibinfo  {journal} {Phys. Lett.}\ }\textbf {\bibinfo {volume}
  {B70}},\ \bibinfo {pages} {433} (\bibinfo {year} {1977})}\BibitemShut
  {NoStop}%
\bibitem [{\citenamefont {Cheng}\ and\ \citenamefont
  {Li}(1980)}]{Cheng:1980qt}%
  \BibitemOpen
  \bibfield  {author} {\bibinfo {author} {\bibfnamefont {T.~P.}\ \bibnamefont
  {Cheng}}\ and\ \bibinfo {author} {\bibfnamefont {L.-F.}\ \bibnamefont {Li}},\
  }\href {\doibase 10.1103/PhysRevD.22.2860} {\bibfield  {journal} {\bibinfo
  {journal} {Phys. Rev.}\ }\textbf {\bibinfo {volume} {D22}},\ \bibinfo {pages}
  {2860} (\bibinfo {year} {1980})}\BibitemShut {NoStop}%
\bibitem [{\citenamefont {Lazarides}\ \emph {et~al.}(1981)\citenamefont
  {Lazarides}, \citenamefont {Shafi},\ and\ \citenamefont
  {Wetterich}}]{Lazarides:1980nt}%
  \BibitemOpen
  \bibfield  {author} {\bibinfo {author} {\bibfnamefont {G.}~\bibnamefont
  {Lazarides}}, \bibinfo {author} {\bibfnamefont {Q.}~\bibnamefont {Shafi}}, \
  and\ \bibinfo {author} {\bibfnamefont {C.}~\bibnamefont {Wetterich}},\ }\href
  {\doibase 10.1016/0550-3213(81)90354-0} {\bibfield  {journal} {\bibinfo
  {journal} {Nucl. Phys.}\ }\textbf {\bibinfo {volume} {B181}},\ \bibinfo
  {pages} {287} (\bibinfo {year} {1981})}\BibitemShut {NoStop}%
\bibitem [{\citenamefont {Mohapatra}\ and\ \citenamefont
  {Senjanovic}(1981)}]{Mohapatra:1980yp}%
  \BibitemOpen
  \bibfield  {author} {\bibinfo {author} {\bibfnamefont {R.~N.}\ \bibnamefont
  {Mohapatra}}\ and\ \bibinfo {author} {\bibfnamefont {G.}~\bibnamefont
  {Senjanovic}},\ }\href {\doibase 10.1103/PhysRevD.23.165} {\bibfield
  {journal} {\bibinfo  {journal} {Phys. Rev.}\ }\textbf {\bibinfo {volume}
  {D23}},\ \bibinfo {pages} {165} (\bibinfo {year} {1981})}\BibitemShut
  {NoStop}%
\bibitem [{\citenamefont {Schechter}\ and\ \citenamefont
  {Valle}(1980)}]{Schechter:1980gr}%
  \BibitemOpen
  \bibfield  {author} {\bibinfo {author} {\bibfnamefont {J.}~\bibnamefont
  {Schechter}}\ and\ \bibinfo {author} {\bibfnamefont {J.~W.~F.}\ \bibnamefont
  {Valle}},\ }\href {\doibase 10.1103/PhysRevD.22.2227} {\bibfield  {journal}
  {\bibinfo  {journal} {Phys. Rev.}\ }\textbf {\bibinfo {volume} {D22}},\
  \bibinfo {pages} {2227} (\bibinfo {year} {1980})}\BibitemShut {NoStop}%
\bibitem [{\citenamefont {Schechter}\ and\ \citenamefont
  {Valle}(1982)}]{Schechter:1981cv}%
  \BibitemOpen
  \bibfield  {author} {\bibinfo {author} {\bibfnamefont {J.}~\bibnamefont
  {Schechter}}\ and\ \bibinfo {author} {\bibfnamefont {J.~W.~F.}\ \bibnamefont
  {Valle}},\ }\href {\doibase 10.1103/PhysRevD.25.774} {\bibfield  {journal}
  {\bibinfo  {journal} {Phys. Rev.}\ }\textbf {\bibinfo {volume} {D25}},\
  \bibinfo {pages} {774} (\bibinfo {year} {1982})}\BibitemShut {NoStop}%
\bibitem [{\citenamefont {Pati}\ and\ \citenamefont
  {Salam}(1974)}]{Pati:1974yy}%
  \BibitemOpen
  \bibfield  {author} {\bibinfo {author} {\bibfnamefont {J.~C.}\ \bibnamefont
  {Pati}}\ and\ \bibinfo {author} {\bibfnamefont {A.}~\bibnamefont {Salam}},\
  }\href {\doibase 10.1103/PhysRevD.10.275, 10.1103/PhysRevD.11.703.2}
  {\bibfield  {journal} {\bibinfo  {journal} {Phys. Rev.}\ }\textbf {\bibinfo
  {volume} {D10}},\ \bibinfo {pages} {275} (\bibinfo {year} {1974})},\ \bibinfo
  {note} {[Erratum: Phys. Rev.D11,703(1975)]}\BibitemShut {NoStop}%
\bibitem [{\citenamefont {Mohapatra}\ and\ \citenamefont
  {Pati}(1975)}]{Mohapatra:1974gc}%
  \BibitemOpen
  \bibfield  {author} {\bibinfo {author} {\bibfnamefont {R.~N.}\ \bibnamefont
  {Mohapatra}}\ and\ \bibinfo {author} {\bibfnamefont {J.~C.}\ \bibnamefont
  {Pati}},\ }\href {\doibase 10.1103/PhysRevD.11.2558} {\bibfield  {journal}
  {\bibinfo  {journal} {Phys. Rev.}\ }\textbf {\bibinfo {volume} {D11}},\
  \bibinfo {pages} {2558} (\bibinfo {year} {1975})}\BibitemShut {NoStop}%
\bibitem [{\citenamefont {Senjanovic}\ and\ \citenamefont
  {Mohapatra}(1975)}]{Senjanovic:1975rk}%
  \BibitemOpen
  \bibfield  {author} {\bibinfo {author} {\bibfnamefont {G.}~\bibnamefont
  {Senjanovic}}\ and\ \bibinfo {author} {\bibfnamefont {R.~N.}\ \bibnamefont
  {Mohapatra}},\ }\href {\doibase 10.1103/PhysRevD.12.1502} {\bibfield
  {journal} {\bibinfo  {journal} {Phys. Rev.}\ }\textbf {\bibinfo {volume}
  {D12}},\ \bibinfo {pages} {1502} (\bibinfo {year} {1975})}\BibitemShut
  {NoStop}%
\bibitem [{\citenamefont {Jenkins}(1987)}]{Jenkins:1987ue}%
  \BibitemOpen
  \bibfield  {author} {\bibinfo {author} {\bibfnamefont {E.~E.}\ \bibnamefont
  {Jenkins}},\ }\href {\doibase 10.1016/0370-2693(87)91172-5} {\bibfield
  {journal} {\bibinfo  {journal} {Phys. Lett.}\ }\textbf {\bibinfo {volume}
  {B192}},\ \bibinfo {pages} {219} (\bibinfo {year} {1987})}\BibitemShut
  {NoStop}%
\bibitem [{\citenamefont {He}\ \emph {et~al.}(1991{\natexlab{a}})\citenamefont
  {He}, \citenamefont {Joshi}, \citenamefont {Lew},\ and\ \citenamefont
  {Volkas}}]{He:1990pn}%
  \BibitemOpen
  \bibfield  {author} {\bibinfo {author} {\bibfnamefont {X.~G.}\ \bibnamefont
  {He}}, \bibinfo {author} {\bibfnamefont {G.~C.}\ \bibnamefont {Joshi}},
  \bibinfo {author} {\bibfnamefont {H.}~\bibnamefont {Lew}}, \ and\ \bibinfo
  {author} {\bibfnamefont {R.~R.}\ \bibnamefont {Volkas}},\ }\href {\doibase
  10.1103/PhysRevD.43.R22} {\bibfield  {journal} {\bibinfo  {journal} {Phys.
  Rev.}\ }\textbf {\bibinfo {volume} {D43}},\ \bibinfo {pages} {22} (\bibinfo
  {year} {1991}{\natexlab{a}})}\BibitemShut {NoStop}%
\bibitem [{\citenamefont {Buchmuller}\ \emph {et~al.}(1991)\citenamefont
  {Buchmuller}, \citenamefont {Greub},\ and\ \citenamefont
  {Minkowski}}]{Buchmuller:1991ce}%
  \BibitemOpen
  \bibfield  {author} {\bibinfo {author} {\bibfnamefont {W.}~\bibnamefont
  {Buchmuller}}, \bibinfo {author} {\bibfnamefont {C.}~\bibnamefont {Greub}}, \
  and\ \bibinfo {author} {\bibfnamefont {P.}~\bibnamefont {Minkowski}},\ }\href
  {\doibase 10.1016/0370-2693(91)90952-M} {\bibfield  {journal} {\bibinfo
  {journal} {Phys. Lett.}\ }\textbf {\bibinfo {volume} {B267}},\ \bibinfo
  {pages} {395} (\bibinfo {year} {1991})}\BibitemShut {NoStop}%
\bibitem [{\citenamefont {He}\ \emph {et~al.}(1991{\natexlab{b}})\citenamefont
  {He}, \citenamefont {Joshi}, \citenamefont {Lew},\ and\ \citenamefont
  {Volkas}}]{He:1991qd}%
  \BibitemOpen
  \bibfield  {author} {\bibinfo {author} {\bibfnamefont {X.-G.}\ \bibnamefont
  {He}}, \bibinfo {author} {\bibfnamefont {G.~C.}\ \bibnamefont {Joshi}},
  \bibinfo {author} {\bibfnamefont {H.}~\bibnamefont {Lew}}, \ and\ \bibinfo
  {author} {\bibfnamefont {R.~R.}\ \bibnamefont {Volkas}},\ }\href {\doibase
  10.1103/PhysRevD.44.2118} {\bibfield  {journal} {\bibinfo  {journal} {Phys.
  Rev.}\ }\textbf {\bibinfo {volume} {D44}},\ \bibinfo {pages} {2118} (\bibinfo
  {year} {1991}{\natexlab{b}})}\BibitemShut {NoStop}%
\bibitem [{\citenamefont {Foot}\ \emph {et~al.}(1994)\citenamefont {Foot},
  \citenamefont {He}, \citenamefont {Lew},\ and\ \citenamefont
  {Volkas}}]{Foot:1994vd}%
  \BibitemOpen
  \bibfield  {author} {\bibinfo {author} {\bibfnamefont {R.}~\bibnamefont
  {Foot}}, \bibinfo {author} {\bibfnamefont {X.~G.}\ \bibnamefont {He}},
  \bibinfo {author} {\bibfnamefont {H.}~\bibnamefont {Lew}}, \ and\ \bibinfo
  {author} {\bibfnamefont {R.~R.}\ \bibnamefont {Volkas}},\ }\href {\doibase
  10.1103/PhysRevD.50.4571} {\bibfield  {journal} {\bibinfo  {journal} {Phys.
  Rev.}\ }\textbf {\bibinfo {volume} {D50}},\ \bibinfo {pages} {4571} (\bibinfo
  {year} {1994})},\ \Eprint {http://arxiv.org/abs/hep-ph/9401250}
  {arXiv:hep-ph/9401250 [hep-ph]} \BibitemShut {NoStop}%
\bibitem [{\citenamefont {Emam}\ and\ \citenamefont
  {Khalil}(2007)}]{Emam:2007dy}%
  \BibitemOpen
  \bibfield  {author} {\bibinfo {author} {\bibfnamefont {W.}~\bibnamefont
  {Emam}}\ and\ \bibinfo {author} {\bibfnamefont {S.}~\bibnamefont {Khalil}},\
  }\href {\doibase 10.1140/epjc/s10052-007-0411-7} {\bibfield  {journal}
  {\bibinfo  {journal} {Eur. Phys. J.}\ }\textbf {\bibinfo {volume} {C52}},\
  \bibinfo {pages} {625} (\bibinfo {year} {2007})},\ \Eprint
  {http://arxiv.org/abs/0704.1395} {arXiv:0704.1395 [hep-ph]} \BibitemShut
  {NoStop}%
\bibitem [{\citenamefont {Basso}\ \emph {et~al.}(2009)\citenamefont {Basso},
  \citenamefont {Belyaev}, \citenamefont {Moretti},\ and\ \citenamefont
  {Shepherd-Themistocleous}}]{Basso:2008iv}%
  \BibitemOpen
  \bibfield  {author} {\bibinfo {author} {\bibfnamefont {L.}~\bibnamefont
  {Basso}}, \bibinfo {author} {\bibfnamefont {A.}~\bibnamefont {Belyaev}},
  \bibinfo {author} {\bibfnamefont {S.}~\bibnamefont {Moretti}}, \ and\
  \bibinfo {author} {\bibfnamefont {C.~H.}\ \bibnamefont
  {Shepherd-Themistocleous}},\ }\href {\doibase 10.1103/PhysRevD.80.055030}
  {\bibfield  {journal} {\bibinfo  {journal} {Phys. Rev.}\ }\textbf {\bibinfo
  {volume} {D80}},\ \bibinfo {pages} {055030} (\bibinfo {year} {2009})},\
  \Eprint {http://arxiv.org/abs/0812.4313} {arXiv:0812.4313 [hep-ph]}
  \BibitemShut {NoStop}%
\bibitem [{\citenamefont {Sevda}\ \emph
  {et~al.}(2017{\natexlab{a}})\citenamefont {Sevda}, \citenamefont {Deniz},
  \citenamefont {Kerman}, \citenamefont {Singh}, \citenamefont {Wong},\ and\
  \citenamefont {Zeyrek}}]{Sevda:2016otj}%
  \BibitemOpen
  \bibfield  {author} {\bibinfo {author} {\bibfnamefont {B.}~\bibnamefont
  {Sevda}}, \bibinfo {author} {\bibfnamefont {M.}~\bibnamefont {Deniz}},
  \bibinfo {author} {\bibfnamefont {S.}~\bibnamefont {Kerman}}, \bibinfo
  {author} {\bibfnamefont {L.}~\bibnamefont {Singh}}, \bibinfo {author}
  {\bibfnamefont {H.~T.}\ \bibnamefont {Wong}}, \ and\ \bibinfo {author}
  {\bibfnamefont {M.}~\bibnamefont {Zeyrek}},\ }\href {\doibase
  10.1103/PhysRevD.95.033008} {\bibfield  {journal} {\bibinfo  {journal} {Phys.
  Rev.}\ }\textbf {\bibinfo {volume} {D95}},\ \bibinfo {pages} {033008}
  (\bibinfo {year} {2017}{\natexlab{a}})},\ \Eprint
  {http://arxiv.org/abs/1611.07259} {arXiv:1611.07259 [hep-ex]} \BibitemShut
  {NoStop}%
\bibitem [{\citenamefont {Deniz}\ \emph
  {et~al.}(2010{\natexlab{a}})\citenamefont {Deniz} \emph
  {et~al.}}]{Deniz:2009mu}%
  \BibitemOpen
  \bibfield  {author} {\bibinfo {author} {\bibfnamefont {M.}~\bibnamefont
  {Deniz}} \emph {et~al.} (\bibinfo {collaboration} {TEXONO}),\ }\href
  {\doibase 10.1103/PhysRevD.81.072001} {\bibfield  {journal} {\bibinfo
  {journal} {Phys. Rev.}\ }\textbf {\bibinfo {volume} {D81}},\ \bibinfo {pages}
  {072001} (\bibinfo {year} {2010}{\natexlab{a}})},\ \Eprint
  {http://arxiv.org/abs/0911.1597} {arXiv:0911.1597 [hep-ex]} \BibitemShut
  {NoStop}%
\bibitem [{\citenamefont {Forero}\ and\ \citenamefont
  {Guzzo}(2011)}]{Forero:2011zz}%
  \BibitemOpen
  \bibfield  {author} {\bibinfo {author} {\bibfnamefont {D.~V.}\ \bibnamefont
  {Forero}}\ and\ \bibinfo {author} {\bibfnamefont {M.~M.}\ \bibnamefont
  {Guzzo}},\ }\href {\doibase 10.1103/PhysRevD.84.013002} {\bibfield  {journal}
  {\bibinfo  {journal} {Phys. Rev.}\ }\textbf {\bibinfo {volume} {D84}},\
  \bibinfo {pages} {013002} (\bibinfo {year} {2011})}\BibitemShut {NoStop}%
\bibitem [{\citenamefont {Kaneta}\ and\ \citenamefont
  {Shimomura}(2017)}]{Kaneta:2016uyt}%
  \BibitemOpen
  \bibfield  {author} {\bibinfo {author} {\bibfnamefont {Y.}~\bibnamefont
  {Kaneta}}\ and\ \bibinfo {author} {\bibfnamefont {T.}~\bibnamefont
  {Shimomura}},\ }\href {\doibase 10.1093/ptep/ptx050} {\bibfield  {journal}
  {\bibinfo  {journal} {PTEP}\ }\textbf {\bibinfo {volume} {2017}},\ \bibinfo
  {pages} {053B04} (\bibinfo {year} {2017})},\ \Eprint
  {http://arxiv.org/abs/1701.00156} {arXiv:1701.00156 [hep-ph]} \BibitemShut
  {NoStop}%
\bibitem [{\citenamefont {Rodejohann}\ \emph {et~al.}(2017)\citenamefont
  {Rodejohann}, \citenamefont {Xu},\ and\ \citenamefont
  {Yaguna}}]{Rodejohann:2017vup}%
  \BibitemOpen
  \bibfield  {author} {\bibinfo {author} {\bibfnamefont {W.}~\bibnamefont
  {Rodejohann}}, \bibinfo {author} {\bibfnamefont {X.-J.}\ \bibnamefont {Xu}},
  \ and\ \bibinfo {author} {\bibfnamefont {C.~E.}\ \bibnamefont {Yaguna}},\
  }\href {\doibase 10.1007/JHEP05(2017)024} {\bibfield  {journal} {\bibinfo
  {journal} {JHEP}\ }\textbf {\bibinfo {volume} {05}},\ \bibinfo {pages} {024}
  (\bibinfo {year} {2017})},\ \Eprint {http://arxiv.org/abs/1702.05721}
  {arXiv:1702.05721 [hep-ph]} \BibitemShut {NoStop}%
\bibitem [{\citenamefont {Lindner}\ \emph {et~al.}(2018)\citenamefont
  {Lindner}, \citenamefont {Queiroz}, \citenamefont {Rodejohann},\ and\
  \citenamefont {Xu}}]{Lindner:2018kjo}%
  \BibitemOpen
  \bibfield  {author} {\bibinfo {author} {\bibfnamefont {M.}~\bibnamefont
  {Lindner}}, \bibinfo {author} {\bibfnamefont {F.~S.}\ \bibnamefont
  {Queiroz}}, \bibinfo {author} {\bibfnamefont {W.}~\bibnamefont {Rodejohann}},
  \ and\ \bibinfo {author} {\bibfnamefont {X.-J.}\ \bibnamefont {Xu}},\ }\href
  {\doibase 10.1007/JHEP05(2018)098} {\bibfield  {journal} {\bibinfo  {journal}
  {JHEP}\ }\textbf {\bibinfo {volume} {05}},\ \bibinfo {pages} {098} (\bibinfo
  {year} {2018})},\ \Eprint {http://arxiv.org/abs/1803.00060} {arXiv:1803.00060
  [hep-ph]} \BibitemShut {NoStop}%
\bibitem [{\citenamefont {Bischer}\ and\ \citenamefont
  {Rodejohann}(2018)}]{Bischer:2018zcz}%
  \BibitemOpen
  \bibfield  {author} {\bibinfo {author} {\bibfnamefont {I.}~\bibnamefont
  {Bischer}}\ and\ \bibinfo {author} {\bibfnamefont {W.}~\bibnamefont
  {Rodejohann}},\ }\href@noop {} {\  (\bibinfo {year} {2018})},\ \Eprint
  {http://arxiv.org/abs/1810.02220} {arXiv:1810.02220 [hep-ph]} \BibitemShut
  {NoStop}%
\bibitem [{\citenamefont {Arguelles}\ \emph {et~al.}(2018)\citenamefont
  {Arguelles}, \citenamefont {Hostert},\ and\ \citenamefont
  {Tsai}}]{Arguelles:2018mtc}%
  \BibitemOpen
  \bibfield  {author} {\bibinfo {author} {\bibfnamefont {C.~A.}\ \bibnamefont
  {Arguelles}}, \bibinfo {author} {\bibfnamefont {M.}~\bibnamefont {Hostert}},
  \ and\ \bibinfo {author} {\bibfnamefont {Y.-D.}\ \bibnamefont {Tsai}},\
  }\href@noop {} {\  (\bibinfo {year} {2018})},\ \Eprint
  {http://arxiv.org/abs/1812.08768} {arXiv:1812.08768 [hep-ph]} \BibitemShut
  {NoStop}%
\bibitem [{\citenamefont {Sevda}\ \emph
  {et~al.}(2017{\natexlab{b}})\citenamefont {Sevda} \emph
  {et~al.}}]{Deniz:2017zok}%
  \BibitemOpen
  \bibfield  {author} {\bibinfo {author} {\bibfnamefont {B.}~\bibnamefont
  {Sevda}} \emph {et~al.},\ }\href {\doibase 10.1103/PhysRevD.96.035017}
  {\bibfield  {journal} {\bibinfo  {journal} {Phys. Rev.}\ }\textbf {\bibinfo
  {volume} {D96}},\ \bibinfo {pages} {035017} (\bibinfo {year}
  {2017}{\natexlab{b}})},\ \Eprint {http://arxiv.org/abs/1702.02353}
  {arXiv:1702.02353 [hep-ph]} \BibitemShut {NoStop}%
\bibitem [{\citenamefont {Kouzakov}\ and\ \citenamefont
  {Studenikin}(2017)}]{Kouzakov:2017hbc}%
  \BibitemOpen
  \bibfield  {author} {\bibinfo {author} {\bibfnamefont {K.~A.}\ \bibnamefont
  {Kouzakov}}\ and\ \bibinfo {author} {\bibfnamefont {A.~I.}\ \bibnamefont
  {Studenikin}},\ }\href {\doibase 10.1103/PhysRevD.95.055013,
  10.1103/PhysRevD.96.099904} {\bibfield  {journal} {\bibinfo  {journal} {Phys.
  Rev.}\ }\textbf {\bibinfo {volume} {D95}},\ \bibinfo {pages} {055013}
  (\bibinfo {year} {2017})},\ \bibinfo {note} {[Erratum: Phys.
  Rev.D96,no.9,099904(2017)]},\ \Eprint {http://arxiv.org/abs/1703.00401}
  {arXiv:1703.00401 [hep-ph]} \BibitemShut {NoStop}%
\bibitem [{\citenamefont {Khan}\ and\ \citenamefont
  {McKay}(2017)}]{Khan:2017oxw}%
  \BibitemOpen
  \bibfield  {author} {\bibinfo {author} {\bibfnamefont {A.~N.}\ \bibnamefont
  {Khan}}\ and\ \bibinfo {author} {\bibfnamefont {D.~W.}\ \bibnamefont
  {McKay}},\ }\href {\doibase 10.1007/JHEP07(2017)143} {\bibfield  {journal}
  {\bibinfo  {journal} {JHEP}\ }\textbf {\bibinfo {volume} {07}},\ \bibinfo
  {pages} {143} (\bibinfo {year} {2017})},\ \Eprint
  {http://arxiv.org/abs/1704.06222} {arXiv:1704.06222 [hep-ph]} \BibitemShut
  {NoStop}%
\bibitem [{\citenamefont {Khan}(2016)}]{Khan:2016uon}%
  \BibitemOpen
  \bibfield  {author} {\bibinfo {author} {\bibfnamefont {A.~N.}\ \bibnamefont
  {Khan}},\ }\href {\doibase 10.1103/PhysRevD.93.093019} {\bibfield  {journal}
  {\bibinfo  {journal} {Phys. Rev.}\ }\textbf {\bibinfo {volume} {D93}},\
  \bibinfo {pages} {093019} (\bibinfo {year} {2016})},\ \Eprint
  {http://arxiv.org/abs/1605.09284} {arXiv:1605.09284 [hep-ph]} \BibitemShut
  {NoStop}%
\bibitem [{\citenamefont {Khan}(2019)}]{Khan:2017djo}%
  \BibitemOpen
  \bibfield  {author} {\bibinfo {author} {\bibfnamefont {A.~N.}\ \bibnamefont
  {Khan}},\ }\href {\doibase 10.1088/1361-6471/ab0057} {\bibfield  {journal}
  {\bibinfo  {journal} {J. Phys.}\ }\textbf {\bibinfo {volume} {G46}},\
  \bibinfo {pages} {035005} (\bibinfo {year} {2019})},\ \Eprint
  {http://arxiv.org/abs/1709.02930} {arXiv:1709.02930 [hep-ph]} \BibitemShut
  {NoStop}%
\bibitem [{\citenamefont {Babu}\ \emph {et~al.}(2017)\citenamefont {Babu},
  \citenamefont {Friedland}, \citenamefont {Machado},\ and\ \citenamefont
  {Mocioiu}}]{Babu:2017olk}%
  \BibitemOpen
  \bibfield  {author} {\bibinfo {author} {\bibfnamefont {K.~S.}\ \bibnamefont
  {Babu}}, \bibinfo {author} {\bibfnamefont {A.}~\bibnamefont {Friedland}},
  \bibinfo {author} {\bibfnamefont {P.~A.~N.}\ \bibnamefont {Machado}}, \ and\
  \bibinfo {author} {\bibfnamefont {I.}~\bibnamefont {Mocioiu}},\ }\href
  {\doibase 10.1007/JHEP12(2017)096} {\bibfield  {journal} {\bibinfo  {journal}
  {JHEP}\ }\textbf {\bibinfo {volume} {12}},\ \bibinfo {pages} {096} (\bibinfo
  {year} {2017})},\ \Eprint {http://arxiv.org/abs/1705.01822} {arXiv:1705.01822
  [hep-ph]} \BibitemShut {NoStop}%
\bibitem [{\citenamefont {Campos}\ \emph {et~al.}(2017)\citenamefont {Campos},
  \citenamefont {Cogollo}, \citenamefont {Lindner}, \citenamefont {Melo},
  \citenamefont {Queiroz},\ and\ \citenamefont {Rodejohann}}]{Campos:2017dgc}%
  \BibitemOpen
  \bibfield  {author} {\bibinfo {author} {\bibfnamefont {M.~D.}\ \bibnamefont
  {Campos}}, \bibinfo {author} {\bibfnamefont {D.}~\bibnamefont {Cogollo}},
  \bibinfo {author} {\bibfnamefont {M.}~\bibnamefont {Lindner}}, \bibinfo
  {author} {\bibfnamefont {T.}~\bibnamefont {Melo}}, \bibinfo {author}
  {\bibfnamefont {F.~S.}\ \bibnamefont {Queiroz}}, \ and\ \bibinfo {author}
  {\bibfnamefont {W.}~\bibnamefont {Rodejohann}},\ }\href {\doibase
  10.1007/JHEP08(2017)092} {\bibfield  {journal} {\bibinfo  {journal} {JHEP}\
  }\textbf {\bibinfo {volume} {08}},\ \bibinfo {pages} {092} (\bibinfo {year}
  {2017})},\ \Eprint {http://arxiv.org/abs/1705.05388} {arXiv:1705.05388
  [hep-ph]} \BibitemShut {NoStop}%
\bibitem [{\citenamefont {Bauer}\ \emph {et~al.}(2018)\citenamefont {Bauer},
  \citenamefont {Foldenauer},\ and\ \citenamefont {Jaeckel}}]{Bauer:2018onh}%
  \BibitemOpen
  \bibfield  {author} {\bibinfo {author} {\bibfnamefont {M.}~\bibnamefont
  {Bauer}}, \bibinfo {author} {\bibfnamefont {P.}~\bibnamefont {Foldenauer}}, \
  and\ \bibinfo {author} {\bibfnamefont {J.}~\bibnamefont {Jaeckel}},\ }\href
  {\doibase 10.1007/JHEP07(2018)094} {\bibfield  {journal} {\bibinfo  {journal}
  {JHEP}\ }\textbf {\bibinfo {volume} {07}},\ \bibinfo {pages} {094} (\bibinfo
  {year} {2018})},\ \Eprint {http://arxiv.org/abs/1803.05466} {arXiv:1803.05466
  [hep-ph]} \BibitemShut {NoStop}%
\bibitem [{\citenamefont {Vilain}\ \emph {et~al.}(1993)\citenamefont {Vilain}
  \emph {et~al.}}]{Vilain:1993kd}%
  \BibitemOpen
  \bibfield  {author} {\bibinfo {author} {\bibfnamefont {P.}~\bibnamefont
  {Vilain}} \emph {et~al.} (\bibinfo {collaboration} {CHARM-II}),\ }\href
  {\doibase 10.1016/0370-2693(93)90408-A} {\bibfield  {journal} {\bibinfo
  {journal} {Phys. Lett.}\ }\textbf {\bibinfo {volume} {B302}},\ \bibinfo
  {pages} {351} (\bibinfo {year} {1993})}\BibitemShut {NoStop}%
\bibitem [{\citenamefont {Vilain}\ \emph {et~al.}(1994)\citenamefont {Vilain}
  \emph {et~al.}}]{Vilain:1994qy}%
  \BibitemOpen
  \bibfield  {author} {\bibinfo {author} {\bibfnamefont {P.}~\bibnamefont
  {Vilain}} \emph {et~al.} (\bibinfo {collaboration} {CHARM-II}),\ }\href
  {\doibase 10.1016/0370-2693(94)91421-4} {\bibfield  {journal} {\bibinfo
  {journal} {Phys. Lett.}\ }\textbf {\bibinfo {volume} {B335}},\ \bibinfo
  {pages} {246} (\bibinfo {year} {1994})}\BibitemShut {NoStop}%
\bibitem [{\citenamefont {Auerbach}\ \emph {et~al.}(2001)\citenamefont
  {Auerbach} \emph {et~al.}}]{Auerbach:2001wg}%
  \BibitemOpen
  \bibfield  {author} {\bibinfo {author} {\bibfnamefont {L.~B.}\ \bibnamefont
  {Auerbach}} \emph {et~al.} (\bibinfo {collaboration} {LSND}),\ }\href
  {\doibase 10.1103/PhysRevD.63.112001} {\bibfield  {journal} {\bibinfo
  {journal} {Phys. Rev.}\ }\textbf {\bibinfo {volume} {D63}},\ \bibinfo {pages}
  {112001} (\bibinfo {year} {2001})},\ \Eprint
  {http://arxiv.org/abs/hep-ex/0101039} {arXiv:hep-ex/0101039 [hep-ex]}
  \BibitemShut {NoStop}%
\bibitem [{\citenamefont {Pospelov}(2011)}]{Pospelov:2011ha}%
  \BibitemOpen
  \bibfield  {author} {\bibinfo {author} {\bibfnamefont {M.}~\bibnamefont
  {Pospelov}},\ }\href {\doibase 10.1103/PhysRevD.84.085008} {\bibfield
  {journal} {\bibinfo  {journal} {Phys. Rev.}\ }\textbf {\bibinfo {volume}
  {D84}},\ \bibinfo {pages} {085008} (\bibinfo {year} {2011})},\ \Eprint
  {http://arxiv.org/abs/1103.3261} {arXiv:1103.3261 [hep-ph]} \BibitemShut
  {NoStop}%
\bibitem [{\citenamefont {Harnik}\ \emph {et~al.}(2012)\citenamefont {Harnik},
  \citenamefont {Kopp},\ and\ \citenamefont {Machado}}]{Harnik:2012ni}%
  \BibitemOpen
  \bibfield  {author} {\bibinfo {author} {\bibfnamefont {R.}~\bibnamefont
  {Harnik}}, \bibinfo {author} {\bibfnamefont {J.}~\bibnamefont {Kopp}}, \ and\
  \bibinfo {author} {\bibfnamefont {P.~A.~N.}\ \bibnamefont {Machado}},\ }\href
  {\doibase 10.1088/1475-7516/2012/07/026} {\bibfield  {journal} {\bibinfo
  {journal} {JCAP}\ }\textbf {\bibinfo {volume} {1207}},\ \bibinfo {pages}
  {026} (\bibinfo {year} {2012})},\ \Eprint {http://arxiv.org/abs/1202.6073}
  {arXiv:1202.6073 [hep-ph]} \BibitemShut {NoStop}%
\bibitem [{\citenamefont {Pospelov}\ and\ \citenamefont
  {Pradler}(2012)}]{Pospelov:2012gm}%
  \BibitemOpen
  \bibfield  {author} {\bibinfo {author} {\bibfnamefont {M.}~\bibnamefont
  {Pospelov}}\ and\ \bibinfo {author} {\bibfnamefont {J.}~\bibnamefont
  {Pradler}},\ }\href {\doibase 10.1103/PhysRevD.85.113016,
  10.1103/PhysRevD.88.039904} {\bibfield  {journal} {\bibinfo  {journal} {Phys.
  Rev.}\ }\textbf {\bibinfo {volume} {D85}},\ \bibinfo {pages} {113016}
  (\bibinfo {year} {2012})},\ \bibinfo {note} {[Erratum: Phys.
  Rev.D88,no.3,039904(2013)]},\ \Eprint {http://arxiv.org/abs/1203.0545}
  {arXiv:1203.0545 [hep-ph]} \BibitemShut {NoStop}%
\bibitem [{\citenamefont {Pospelov}\ and\ \citenamefont
  {Pradler}(2014)}]{Pospelov:2013rha}%
  \BibitemOpen
  \bibfield  {author} {\bibinfo {author} {\bibfnamefont {M.}~\bibnamefont
  {Pospelov}}\ and\ \bibinfo {author} {\bibfnamefont {J.}~\bibnamefont
  {Pradler}},\ }\href {\doibase 10.1103/PhysRevD.89.055012} {\bibfield
  {journal} {\bibinfo  {journal} {Phys. Rev.}\ }\textbf {\bibinfo {volume}
  {D89}},\ \bibinfo {pages} {055012} (\bibinfo {year} {2014})},\ \Eprint
  {http://arxiv.org/abs/1311.5764} {arXiv:1311.5764 [hep-ph]} \BibitemShut
  {NoStop}%
\bibitem [{\citenamefont {Coloma}\ \emph {et~al.}(2014)\citenamefont {Coloma},
  \citenamefont {Huber},\ and\ \citenamefont {Link}}]{Coloma:2014hka}%
  \BibitemOpen
  \bibfield  {author} {\bibinfo {author} {\bibfnamefont {P.}~\bibnamefont
  {Coloma}}, \bibinfo {author} {\bibfnamefont {P.}~\bibnamefont {Huber}}, \
  and\ \bibinfo {author} {\bibfnamefont {J.~M.}\ \bibnamefont {Link}},\ }\href
  {\doibase 10.1007/JHEP11(2014)042} {\bibfield  {journal} {\bibinfo  {journal}
  {JHEP}\ }\textbf {\bibinfo {volume} {11}},\ \bibinfo {pages} {042} (\bibinfo
  {year} {2014})},\ \Eprint {http://arxiv.org/abs/1406.4914} {arXiv:1406.4914
  [hep-ph]} \BibitemShut {NoStop}%
\bibitem [{\citenamefont {Malling}\ \emph {et~al.}(2011)\citenamefont {Malling}
  \emph {et~al.}}]{Malling:2011va}%
  \BibitemOpen
  \bibfield  {author} {\bibinfo {author} {\bibfnamefont {D.~C.}\ \bibnamefont
  {Malling}} \emph {et~al.},\ }\href@noop {} {\enquote {\bibinfo {title}
  {{After LUX: The LZ Program}},}\ } (\bibinfo {year} {2011}),\ \Eprint
  {http://arxiv.org/abs/1110.0103} {arXiv:1110.0103 [astro-ph.IM]} \BibitemShut
  {NoStop}%
\bibitem [{\citenamefont {Akerib}\ \emph {et~al.}(2015)\citenamefont {Akerib}
  \emph {et~al.}}]{Akerib:2015cja}%
  \BibitemOpen
  \bibfield  {author} {\bibinfo {author} {\bibfnamefont {D.~S.}\ \bibnamefont
  {Akerib}} \emph {et~al.} (\bibinfo {collaboration} {LZ}),\ }\href@noop {}
  {\enquote {\bibinfo {title} {{LUX-ZEPLIN (LZ) Conceptual Design Report}},}\ }
  (\bibinfo {year} {2015}),\ \Eprint {http://arxiv.org/abs/1509.02910}
  {arXiv:1509.02910 [physics.ins-det]} \BibitemShut {NoStop}%
\bibitem [{\citenamefont {Mount}\ \emph {et~al.}(2017)\citenamefont {Mount}
  \emph {et~al.}}]{Mount:2017qzi}%
  \BibitemOpen
  \bibfield  {author} {\bibinfo {author} {\bibfnamefont {B.~J.}\ \bibnamefont
  {Mount}} \emph {et~al.},\ }\href@noop {} {\  (\bibinfo {year} {2017})},\
  \Eprint {http://arxiv.org/abs/1703.09144} {arXiv:1703.09144
  [physics.ins-det]} \BibitemShut {NoStop}%
\bibitem [{\citenamefont {Cribier}\ \emph {et~al.}(1996)\citenamefont {Cribier}
  \emph {et~al.}}]{Cribier:1996cq}%
  \BibitemOpen
  \bibfield  {author} {\bibinfo {author} {\bibfnamefont {M.}~\bibnamefont
  {Cribier}} \emph {et~al.},\ }\href {\doibase 10.1016/0168-9002(96)00464-0}
  {\bibfield  {journal} {\bibinfo  {journal} {Nucl. Instrum. Meth.}\ }\textbf
  {\bibinfo {volume} {A378}},\ \bibinfo {pages} {233} (\bibinfo {year}
  {1996})}\BibitemShut {NoStop}%
\bibitem [{\citenamefont {Arcadi}\ \emph {et~al.}(2019)\citenamefont {Arcadi},
  \citenamefont {Lindner}, \citenamefont {Martins},\ and\ \citenamefont
  {Queiroz}}]{Arcadi:2019uif}%
  \BibitemOpen
  \bibfield  {author} {\bibinfo {author} {\bibfnamefont {G.}~\bibnamefont
  {Arcadi}}, \bibinfo {author} {\bibfnamefont {M.}~\bibnamefont {Lindner}},
  \bibinfo {author} {\bibfnamefont {J.}~\bibnamefont {Martins}}, \ and\
  \bibinfo {author} {\bibfnamefont {F.~S.}\ \bibnamefont {Queiroz}},\
  }\href@noop {} {\  (\bibinfo {year} {2019})},\ \Eprint
  {http://arxiv.org/abs/1906.04755} {arXiv:1906.04755 [hep-ph]} \BibitemShut
  {NoStop}%
\bibitem [{\citenamefont {Giunti}\ and\ \citenamefont {Kim}(2007)}]{Giunti}%
  \BibitemOpen
  \bibfield  {author} {\bibinfo {author} {\bibfnamefont {C.}~\bibnamefont
  {Giunti}}\ and\ \bibinfo {author} {\bibfnamefont {C.~W.}\ \bibnamefont
  {Kim}},\ }\href@noop {} {\emph {\bibinfo {title} {{Fundamentals of Neutrino
  Physics and Astrophysics}}}}\ (\bibinfo  {publisher} {Oxford University
  Press},\ \bibinfo {year} {2007})\ pp.\ \bibinfo {pages}
  {136--140}\BibitemShut {NoStop}%
\bibitem [{\citenamefont {Baudis}\ \emph {et~al.}(2014)\citenamefont {Baudis},
  \citenamefont {Ferella}, \citenamefont {Kish}, \citenamefont {Manalaysay},
  \citenamefont {Marrodan~Undagoitia},\ and\ \citenamefont
  {Schumann}}]{Baudis:2013qla}%
  \BibitemOpen
  \bibfield  {author} {\bibinfo {author} {\bibfnamefont {L.}~\bibnamefont
  {Baudis}}, \bibinfo {author} {\bibfnamefont {A.}~\bibnamefont {Ferella}},
  \bibinfo {author} {\bibfnamefont {A.}~\bibnamefont {Kish}}, \bibinfo {author}
  {\bibfnamefont {A.}~\bibnamefont {Manalaysay}}, \bibinfo {author}
  {\bibfnamefont {T.}~\bibnamefont {Marrodan~Undagoitia}}, \ and\ \bibinfo
  {author} {\bibfnamefont {M.}~\bibnamefont {Schumann}},\ }\href {\doibase
  10.1088/1475-7516/2014/01/044} {\bibfield  {journal} {\bibinfo  {journal}
  {JCAP}\ }\textbf {\bibinfo {volume} {1401}},\ \bibinfo {pages} {044}
  (\bibinfo {year} {2014})},\ \Eprint {http://arxiv.org/abs/1309.7024}
  {arXiv:1309.7024 [physics.ins-det]} \BibitemShut {NoStop}%
\bibitem [{\citenamefont {Tanabashi}\ \emph {et~al.}(2018)\citenamefont
  {Tanabashi} \emph {et~al.}}]{Tanabashi:2018oca}%
  \BibitemOpen
  \bibfield  {author} {\bibinfo {author} {\bibfnamefont {M.}~\bibnamefont
  {Tanabashi}} \emph {et~al.} (\bibinfo {collaboration} {Particle Data
  Group}),\ }\href {\doibase 10.1103/PhysRevD.98.030001} {\bibfield  {journal}
  {\bibinfo  {journal} {Phys. Rev.}\ }\textbf {\bibinfo {volume} {D98}},\
  \bibinfo {pages} {030001} (\bibinfo {year} {2018})}\BibitemShut {NoStop}%
\bibitem [{\citenamefont {Erler}\ and\ \citenamefont
  {Ramsey-Musolf}(2005)}]{Erler:2004in}%
  \BibitemOpen
  \bibfield  {author} {\bibinfo {author} {\bibfnamefont {J.}~\bibnamefont
  {Erler}}\ and\ \bibinfo {author} {\bibfnamefont {M.~J.}\ \bibnamefont
  {Ramsey-Musolf}},\ }\href {\doibase 10.1103/PhysRevD.72.073003} {\bibfield
  {journal} {\bibinfo  {journal} {Phys. Rev.}\ }\textbf {\bibinfo {volume}
  {D72}},\ \bibinfo {pages} {073003} (\bibinfo {year} {2005})},\ \Eprint
  {http://arxiv.org/abs/hep-ph/0409169} {arXiv:hep-ph/0409169 [hep-ph]}
  \BibitemShut {NoStop}%
\bibitem [{\citenamefont {Erler}\ and\ \citenamefont
  {Su}(2013)}]{Erler:2013xha}%
  \BibitemOpen
  \bibfield  {author} {\bibinfo {author} {\bibfnamefont {J.}~\bibnamefont
  {Erler}}\ and\ \bibinfo {author} {\bibfnamefont {S.}~\bibnamefont {Su}},\
  }\href {\doibase 10.1016/j.ppnp.2013.03.004} {\bibfield  {journal} {\bibinfo
  {journal} {Prog. Part. Nucl. Phys.}\ }\textbf {\bibinfo {volume} {71}},\
  \bibinfo {pages} {119} (\bibinfo {year} {2013})},\ \Eprint
  {http://arxiv.org/abs/1303.5522} {arXiv:1303.5522 [hep-ph]} \BibitemShut
  {NoStop}%
\bibitem [{\citenamefont {Healey}\ \emph {et~al.}(2013)\citenamefont {Healey},
  \citenamefont {Petrov},\ and\ \citenamefont {Zhuridov}}]{Healey:2013vka}%
  \BibitemOpen
  \bibfield  {author} {\bibinfo {author} {\bibfnamefont {K.~J.}\ \bibnamefont
  {Healey}}, \bibinfo {author} {\bibfnamefont {A.~A.}\ \bibnamefont {Petrov}},
  \ and\ \bibinfo {author} {\bibfnamefont {D.}~\bibnamefont {Zhuridov}},\
  }\href {\doibase 10.1103/PhysRevD.87.117301, 10.1103/PhysRevD.89.059904}
  {\bibfield  {journal} {\bibinfo  {journal} {Phys. Rev.}\ }\textbf {\bibinfo
  {volume} {D87}},\ \bibinfo {pages} {117301} (\bibinfo {year} {2013})},\
  \bibinfo {note} {[Erratum: Phys. Rev.D89,no.5,059904(2014)]},\ \Eprint
  {http://arxiv.org/abs/1305.0584} {arXiv:1305.0584 [hep-ph]} \BibitemShut
  {NoStop}%
\bibitem [{\citenamefont {Lindner}\ \emph {et~al.}(2017)\citenamefont
  {Lindner}, \citenamefont {Rodejohann},\ and\ \citenamefont
  {Xu}}]{Lindner:2016wff}%
  \BibitemOpen
  \bibfield  {author} {\bibinfo {author} {\bibfnamefont {M.}~\bibnamefont
  {Lindner}}, \bibinfo {author} {\bibfnamefont {W.}~\bibnamefont {Rodejohann}},
  \ and\ \bibinfo {author} {\bibfnamefont {X.-J.}\ \bibnamefont {Xu}},\ }\href
  {\doibase 10.1007/JHEP03(2017)097} {\bibfield  {journal} {\bibinfo  {journal}
  {JHEP}\ }\textbf {\bibinfo {volume} {03}},\ \bibinfo {pages} {097} (\bibinfo
  {year} {2017})},\ \Eprint {http://arxiv.org/abs/1612.04150} {arXiv:1612.04150
  [hep-ph]} \BibitemShut {NoStop}%
\bibitem [{\citenamefont {Heurtier}\ and\ \citenamefont
  {Zhang}(2017)}]{Heurtier:2016otg}%
  \BibitemOpen
  \bibfield  {author} {\bibinfo {author} {\bibfnamefont {L.}~\bibnamefont
  {Heurtier}}\ and\ \bibinfo {author} {\bibfnamefont {Y.}~\bibnamefont
  {Zhang}},\ }\href {\doibase 10.1088/1475-7516/2017/02/042} {\bibfield
  {journal} {\bibinfo  {journal} {JCAP}\ }\textbf {\bibinfo {volume} {1702}},\
  \bibinfo {pages} {042} (\bibinfo {year} {2017})},\ \Eprint
  {http://arxiv.org/abs/1609.05882} {arXiv:1609.05882 [hep-ph]} \BibitemShut
  {NoStop}%
\bibitem [{\citenamefont {Papoulias}\ and\ \citenamefont
  {Kosmas}(2018)}]{Kosmas:2017tsq}%
  \BibitemOpen
  \bibfield  {author} {\bibinfo {author} {\bibfnamefont {D.~K.}\ \bibnamefont
  {Papoulias}}\ and\ \bibinfo {author} {\bibfnamefont {T.~S.}\ \bibnamefont
  {Kosmas}},\ }\href {\doibase 10.1103/PhysRevD.97.033003} {\bibfield
  {journal} {\bibinfo  {journal} {Phys. Rev.}\ }\textbf {\bibinfo {volume}
  {D97}},\ \bibinfo {pages} {033003} (\bibinfo {year} {2018})},\ \Eprint
  {http://arxiv.org/abs/1711.09773} {arXiv:1711.09773 [hep-ph]} \BibitemShut
  {NoStop}%
\bibitem [{\citenamefont {Magill}\ and\ \citenamefont
  {Plestid}(2018)}]{Magill:2017mps}%
  \BibitemOpen
  \bibfield  {author} {\bibinfo {author} {\bibfnamefont {G.}~\bibnamefont
  {Magill}}\ and\ \bibinfo {author} {\bibfnamefont {R.}~\bibnamefont
  {Plestid}},\ }\href {\doibase 10.1103/PhysRevD.97.055003} {\bibfield
  {journal} {\bibinfo  {journal} {Phys. Rev.}\ }\textbf {\bibinfo {volume}
  {D97}},\ \bibinfo {pages} {055003} (\bibinfo {year} {2018})},\ \Eprint
  {http://arxiv.org/abs/1710.08431} {arXiv:1710.08431 [hep-ph]} \BibitemShut
  {NoStop}%
\bibitem [{\citenamefont {Farzan}\ \emph {et~al.}(2018)\citenamefont {Farzan},
  \citenamefont {Lindner}, \citenamefont {Rodejohann},\ and\ \citenamefont
  {Xu}}]{Farzan:2018gtr}%
  \BibitemOpen
  \bibfield  {author} {\bibinfo {author} {\bibfnamefont {Y.}~\bibnamefont
  {Farzan}}, \bibinfo {author} {\bibfnamefont {M.}~\bibnamefont {Lindner}},
  \bibinfo {author} {\bibfnamefont {W.}~\bibnamefont {Rodejohann}}, \ and\
  \bibinfo {author} {\bibfnamefont {X.-J.}\ \bibnamefont {Xu}},\ }\href
  {\doibase 10.1007/JHEP05(2018)066} {\bibfield  {journal} {\bibinfo  {journal}
  {JHEP}\ }\textbf {\bibinfo {volume} {05}},\ \bibinfo {pages} {066} (\bibinfo
  {year} {2018})},\ \Eprint {http://arxiv.org/abs/1802.05171} {arXiv:1802.05171
  [hep-ph]} \BibitemShut {NoStop}%
\bibitem [{\citenamefont {Yang}\ and\ \citenamefont
  {Kneller}(2018)}]{Yang:2018yvk}%
  \BibitemOpen
  \bibfield  {author} {\bibinfo {author} {\bibfnamefont {Y.}~\bibnamefont
  {Yang}}\ and\ \bibinfo {author} {\bibfnamefont {J.~P.}\ \bibnamefont
  {Kneller}},\ }\href {\doibase 10.1103/PhysRevD.97.103018} {\bibfield
  {journal} {\bibinfo  {journal} {Phys. Rev.}\ }\textbf {\bibinfo {volume}
  {D97}},\ \bibinfo {pages} {103018} (\bibinfo {year} {2018})},\ \Eprint
  {http://arxiv.org/abs/1803.04504} {arXiv:1803.04504 [astro-ph.HE]}
  \BibitemShut {NoStop}%
\bibitem [{\citenamefont {Aristizabal~Sierra}\ \emph
  {et~al.}(2018)\citenamefont {Aristizabal~Sierra}, \citenamefont {De~Romeri},\
  and\ \citenamefont {Rojas}}]{AristizabalSierra:2018eqm}%
  \BibitemOpen
  \bibfield  {author} {\bibinfo {author} {\bibfnamefont {D.}~\bibnamefont
  {Aristizabal~Sierra}}, \bibinfo {author} {\bibfnamefont {V.}~\bibnamefont
  {De~Romeri}}, \ and\ \bibinfo {author} {\bibfnamefont {N.}~\bibnamefont
  {Rojas}},\ }\href {\doibase 10.1103/PhysRevD.98.075018} {\bibfield  {journal}
  {\bibinfo  {journal} {Phys. Rev.}\ }\textbf {\bibinfo {volume} {D98}},\
  \bibinfo {pages} {075018} (\bibinfo {year} {2018})},\ \Eprint
  {http://arxiv.org/abs/1806.07424} {arXiv:1806.07424 [hep-ph]} \BibitemShut
  {NoStop}%
\bibitem [{\citenamefont {Brdar}\ \emph {et~al.}(2018)\citenamefont {Brdar},
  \citenamefont {Rodejohann},\ and\ \citenamefont {Xu}}]{Brdar:2018qqj}%
  \BibitemOpen
  \bibfield  {author} {\bibinfo {author} {\bibfnamefont {V.}~\bibnamefont
  {Brdar}}, \bibinfo {author} {\bibfnamefont {W.}~\bibnamefont {Rodejohann}}, \
  and\ \bibinfo {author} {\bibfnamefont {X.-J.}\ \bibnamefont {Xu}},\ }\href
  {\doibase 10.1007/JHEP12(2018)024} {\bibfield  {journal} {\bibinfo  {journal}
  {JHEP}\ }\textbf {\bibinfo {volume} {12}},\ \bibinfo {pages} {024} (\bibinfo
  {year} {2018})},\ \Eprint {http://arxiv.org/abs/1810.03626} {arXiv:1810.03626
  [hep-ph]} \BibitemShut {NoStop}%
\bibitem [{\citenamefont {Blaut}\ and\ \citenamefont
  {Sobkow}(2018)}]{Blaut:2018fis}%
  \BibitemOpen
  \bibfield  {author} {\bibinfo {author} {\bibfnamefont {A.}~\bibnamefont
  {Blaut}}\ and\ \bibinfo {author} {\bibfnamefont {W.}~\bibnamefont {Sobkow}},\
  }\href@noop {} {\  (\bibinfo {year} {2018})},\ \Eprint
  {http://arxiv.org/abs/1812.09828} {arXiv:1812.09828 [hep-ph]} \BibitemShut
  {NoStop}%
\bibitem [{\citenamefont {Feldman}\ \emph {et~al.}(2007)\citenamefont
  {Feldman}, \citenamefont {Liu},\ and\ \citenamefont {Nath}}]{Feldman:2007wj}%
  \BibitemOpen
  \bibfield  {author} {\bibinfo {author} {\bibfnamefont {D.}~\bibnamefont
  {Feldman}}, \bibinfo {author} {\bibfnamefont {Z.}~\bibnamefont {Liu}}, \ and\
  \bibinfo {author} {\bibfnamefont {P.}~\bibnamefont {Nath}},\ }\href {\doibase
  10.1103/PhysRevD.75.115001} {\bibfield  {journal} {\bibinfo  {journal} {Phys.
  Rev.}\ }\textbf {\bibinfo {volume} {D75}},\ \bibinfo {pages} {115001}
  (\bibinfo {year} {2007})},\ \Eprint {http://arxiv.org/abs/hep-ph/0702123}
  {arXiv:hep-ph/0702123 [HEP-PH]} \BibitemShut {NoStop}%
\bibitem [{\citenamefont {Dudas}\ \emph {et~al.}(2009)\citenamefont {Dudas},
  \citenamefont {Mambrini}, \citenamefont {Pokorski},\ and\ \citenamefont
  {Romagnoni}}]{Dudas:2009uq}%
  \BibitemOpen
  \bibfield  {author} {\bibinfo {author} {\bibfnamefont {E.}~\bibnamefont
  {Dudas}}, \bibinfo {author} {\bibfnamefont {Y.}~\bibnamefont {Mambrini}},
  \bibinfo {author} {\bibfnamefont {S.}~\bibnamefont {Pokorski}}, \ and\
  \bibinfo {author} {\bibfnamefont {A.}~\bibnamefont {Romagnoni}},\ }\href
  {\doibase 10.1088/1126-6708/2009/08/014} {\bibfield  {journal} {\bibinfo
  {journal} {JHEP}\ }\textbf {\bibinfo {volume} {08}},\ \bibinfo {pages} {014}
  (\bibinfo {year} {2009})},\ \Eprint {http://arxiv.org/abs/0904.1745}
  {arXiv:0904.1745 [hep-ph]} \BibitemShut {NoStop}%
\bibitem [{\citenamefont {Buckley}\ \emph {et~al.}(2011)\citenamefont
  {Buckley}, \citenamefont {Hooper}, \citenamefont {Kopp},\ and\ \citenamefont
  {Neil}}]{Buckley:2011vc}%
  \BibitemOpen
  \bibfield  {author} {\bibinfo {author} {\bibfnamefont {M.~R.}\ \bibnamefont
  {Buckley}}, \bibinfo {author} {\bibfnamefont {D.}~\bibnamefont {Hooper}},
  \bibinfo {author} {\bibfnamefont {J.}~\bibnamefont {Kopp}}, \ and\ \bibinfo
  {author} {\bibfnamefont {E.}~\bibnamefont {Neil}},\ }\href {\doibase
  10.1103/PhysRevD.83.115013} {\bibfield  {journal} {\bibinfo  {journal} {Phys.
  Rev.}\ }\textbf {\bibinfo {volume} {D83}},\ \bibinfo {pages} {115013}
  (\bibinfo {year} {2011})},\ \Eprint {http://arxiv.org/abs/1103.6035}
  {arXiv:1103.6035 [hep-ph]} \BibitemShut {NoStop}%
\bibitem [{\citenamefont {An}\ \emph {et~al.}(2012)\citenamefont {An},
  \citenamefont {Ji},\ and\ \citenamefont {Wang}}]{An:2012va}%
  \BibitemOpen
  \bibfield  {author} {\bibinfo {author} {\bibfnamefont {H.}~\bibnamefont
  {An}}, \bibinfo {author} {\bibfnamefont {X.}~\bibnamefont {Ji}}, \ and\
  \bibinfo {author} {\bibfnamefont {L.-T.}\ \bibnamefont {Wang}},\ }\href
  {\doibase 10.1007/JHEP07(2012)182} {\bibfield  {journal} {\bibinfo  {journal}
  {JHEP}\ }\textbf {\bibinfo {volume} {07}},\ \bibinfo {pages} {182} (\bibinfo
  {year} {2012})},\ \Eprint {http://arxiv.org/abs/1202.2894} {arXiv:1202.2894
  [hep-ph]} \BibitemShut {NoStop}%
\bibitem [{\citenamefont {Arcadi}\ \emph {et~al.}(2014)\citenamefont {Arcadi},
  \citenamefont {Mambrini}, \citenamefont {Tytgat},\ and\ \citenamefont
  {Zaldivar}}]{Arcadi:2013qia}%
  \BibitemOpen
  \bibfield  {author} {\bibinfo {author} {\bibfnamefont {G.}~\bibnamefont
  {Arcadi}}, \bibinfo {author} {\bibfnamefont {Y.}~\bibnamefont {Mambrini}},
  \bibinfo {author} {\bibfnamefont {M.~H.~G.}\ \bibnamefont {Tytgat}}, \ and\
  \bibinfo {author} {\bibfnamefont {B.}~\bibnamefont {Zaldivar}},\ }\href
  {\doibase 10.1007/JHEP03(2014)134} {\bibfield  {journal} {\bibinfo  {journal}
  {JHEP}\ }\textbf {\bibinfo {volume} {03}},\ \bibinfo {pages} {134} (\bibinfo
  {year} {2014})},\ \Eprint {http://arxiv.org/abs/1401.0221} {arXiv:1401.0221
  [hep-ph]} \BibitemShut {NoStop}%
\bibitem [{\citenamefont {Alves}\ \emph {et~al.}(2014)\citenamefont {Alves},
  \citenamefont {Profumo},\ and\ \citenamefont {Queiroz}}]{Alves:2013tqa}%
  \BibitemOpen
  \bibfield  {author} {\bibinfo {author} {\bibfnamefont {A.}~\bibnamefont
  {Alves}}, \bibinfo {author} {\bibfnamefont {S.}~\bibnamefont {Profumo}}, \
  and\ \bibinfo {author} {\bibfnamefont {F.~S.}\ \bibnamefont {Queiroz}},\
  }\href {\doibase 10.1007/JHEP04(2014)063} {\bibfield  {journal} {\bibinfo
  {journal} {JHEP}\ }\textbf {\bibinfo {volume} {04}},\ \bibinfo {pages} {063}
  (\bibinfo {year} {2014})},\ \Eprint {http://arxiv.org/abs/1312.5281}
  {arXiv:1312.5281 [hep-ph]} \BibitemShut {NoStop}%
\bibitem [{\citenamefont {Cline}\ \emph {et~al.}(2014)\citenamefont {Cline},
  \citenamefont {Dupuis}, \citenamefont {Liu},\ and\ \citenamefont
  {Xue}}]{Cline:2014dwa}%
  \BibitemOpen
  \bibfield  {author} {\bibinfo {author} {\bibfnamefont {J.~M.}\ \bibnamefont
  {Cline}}, \bibinfo {author} {\bibfnamefont {G.}~\bibnamefont {Dupuis}},
  \bibinfo {author} {\bibfnamefont {Z.}~\bibnamefont {Liu}}, \ and\ \bibinfo
  {author} {\bibfnamefont {W.}~\bibnamefont {Xue}},\ }\href {\doibase
  10.1007/JHEP08(2014)131} {\bibfield  {journal} {\bibinfo  {journal} {JHEP}\
  }\textbf {\bibinfo {volume} {08}},\ \bibinfo {pages} {131} (\bibinfo {year}
  {2014})},\ \Eprint {http://arxiv.org/abs/1405.7691} {arXiv:1405.7691
  [hep-ph]} \BibitemShut {NoStop}%
\bibitem [{\citenamefont {Ducu}\ \emph {et~al.}(2016)\citenamefont {Ducu},
  \citenamefont {Heurtier},\ and\ \citenamefont {Maurer}}]{Ducu:2015fda}%
  \BibitemOpen
  \bibfield  {author} {\bibinfo {author} {\bibfnamefont {O.}~\bibnamefont
  {Ducu}}, \bibinfo {author} {\bibfnamefont {L.}~\bibnamefont {Heurtier}}, \
  and\ \bibinfo {author} {\bibfnamefont {J.}~\bibnamefont {Maurer}},\ }\href
  {\doibase 10.1007/JHEP03(2016)006} {\bibfield  {journal} {\bibinfo  {journal}
  {JHEP}\ }\textbf {\bibinfo {volume} {03}},\ \bibinfo {pages} {006} (\bibinfo
  {year} {2016})},\ \Eprint {http://arxiv.org/abs/1509.05615} {arXiv:1509.05615
  [hep-ph]} \BibitemShut {NoStop}%
\bibitem [{\citenamefont {Buchmueller}\ \emph {et~al.}(2015)\citenamefont
  {Buchmueller}, \citenamefont {Dolan}, \citenamefont {Malik},\ and\
  \citenamefont {McCabe}}]{Buchmueller:2014yoa}%
  \BibitemOpen
  \bibfield  {author} {\bibinfo {author} {\bibfnamefont {O.}~\bibnamefont
  {Buchmueller}}, \bibinfo {author} {\bibfnamefont {M.~J.}\ \bibnamefont
  {Dolan}}, \bibinfo {author} {\bibfnamefont {S.~A.}\ \bibnamefont {Malik}}, \
  and\ \bibinfo {author} {\bibfnamefont {C.}~\bibnamefont {McCabe}},\ }\href
  {\doibase 10.1007/JHEP01(2015)037} {\bibfield  {journal} {\bibinfo  {journal}
  {JHEP}\ }\textbf {\bibinfo {volume} {01}},\ \bibinfo {pages} {037} (\bibinfo
  {year} {2015})},\ \Eprint {http://arxiv.org/abs/1407.8257} {arXiv:1407.8257
  [hep-ph]} \BibitemShut {NoStop}%
\bibitem [{\citenamefont {Alves}\ \emph {et~al.}(2015)\citenamefont {Alves},
  \citenamefont {Berlin}, \citenamefont {Profumo},\ and\ \citenamefont
  {Queiroz}}]{Alves:2015mua}%
  \BibitemOpen
  \bibfield  {author} {\bibinfo {author} {\bibfnamefont {A.}~\bibnamefont
  {Alves}}, \bibinfo {author} {\bibfnamefont {A.}~\bibnamefont {Berlin}},
  \bibinfo {author} {\bibfnamefont {S.}~\bibnamefont {Profumo}}, \ and\
  \bibinfo {author} {\bibfnamefont {F.~S.}\ \bibnamefont {Queiroz}},\ }\href
  {\doibase 10.1007/JHEP10(2015)076} {\bibfield  {journal} {\bibinfo  {journal}
  {JHEP}\ }\textbf {\bibinfo {volume} {10}},\ \bibinfo {pages} {076} (\bibinfo
  {year} {2015})},\ \Eprint {http://arxiv.org/abs/1506.06767} {arXiv:1506.06767
  [hep-ph]} \BibitemShut {NoStop}%
\bibitem [{\citenamefont {Okada}\ and\ \citenamefont
  {Okada}(2016)}]{Okada:2016gsh}%
  \BibitemOpen
  \bibfield  {author} {\bibinfo {author} {\bibfnamefont {N.}~\bibnamefont
  {Okada}}\ and\ \bibinfo {author} {\bibfnamefont {S.}~\bibnamefont {Okada}},\
  }\href {\doibase 10.1103/PhysRevD.93.075003} {\bibfield  {journal} {\bibinfo
  {journal} {Phys. Rev.}\ }\textbf {\bibinfo {volume} {D93}},\ \bibinfo {pages}
  {075003} (\bibinfo {year} {2016})},\ \Eprint
  {http://arxiv.org/abs/1601.07526} {arXiv:1601.07526 [hep-ph]} \BibitemShut
  {NoStop}%
\bibitem [{\citenamefont {Ge}\ and\ \citenamefont
  {Shoemaker}(2018)}]{Ge:2017mcq}%
  \BibitemOpen
  \bibfield  {author} {\bibinfo {author} {\bibfnamefont {S.-F.}\ \bibnamefont
  {Ge}}\ and\ \bibinfo {author} {\bibfnamefont {I.~M.}\ \bibnamefont
  {Shoemaker}},\ }\href {\doibase 10.1007/JHEP11(2018)066} {\bibfield
  {journal} {\bibinfo  {journal} {JHEP}\ }\textbf {\bibinfo {volume} {11}},\
  \bibinfo {pages} {066} (\bibinfo {year} {2018})},\ \Eprint
  {http://arxiv.org/abs/1710.10889} {arXiv:1710.10889 [hep-ph]} \BibitemShut
  {NoStop}%
\bibitem [{\citenamefont {Heeck}\ \emph {et~al.}(2018)\citenamefont {Heeck},
  \citenamefont {Lindner}, \citenamefont {Rodejohann},\ and\ \citenamefont
  {Vogl}}]{Heeck:2018nzc}%
  \BibitemOpen
  \bibfield  {author} {\bibinfo {author} {\bibfnamefont {J.}~\bibnamefont
  {Heeck}}, \bibinfo {author} {\bibfnamefont {M.}~\bibnamefont {Lindner}},
  \bibinfo {author} {\bibfnamefont {W.}~\bibnamefont {Rodejohann}}, \ and\
  \bibinfo {author} {\bibfnamefont {S.}~\bibnamefont {Vogl}},\ }\href@noop {}
  {\  (\bibinfo {year} {2018})},\ \Eprint {http://arxiv.org/abs/1812.04067}
  {arXiv:1812.04067 [hep-ph]} \BibitemShut {NoStop}%
\bibitem [{\citenamefont {Bischer}\ \emph {et~al.}(2018)\citenamefont
  {Bischer}, \citenamefont {Rodejohann},\ and\ \citenamefont
  {Xu}}]{Bischer:2018zbd}%
  \BibitemOpen
  \bibfield  {author} {\bibinfo {author} {\bibfnamefont {I.}~\bibnamefont
  {Bischer}}, \bibinfo {author} {\bibfnamefont {W.}~\bibnamefont {Rodejohann}},
  \ and\ \bibinfo {author} {\bibfnamefont {X.-J.}\ \bibnamefont {Xu}},\ }\href
  {\doibase 10.1007/JHEP10(2018)096} {\bibfield  {journal} {\bibinfo  {journal}
  {JHEP}\ }\textbf {\bibinfo {volume} {10}},\ \bibinfo {pages} {096} (\bibinfo
  {year} {2018})},\ \Eprint {http://arxiv.org/abs/1807.08102} {arXiv:1807.08102
  [hep-ph]} \BibitemShut {NoStop}%
\bibitem [{\citenamefont {Deniz}\ \emph
  {et~al.}(2010{\natexlab{b}})\citenamefont {Deniz} \emph
  {et~al.}}]{Deniz:2010mp}%
  \BibitemOpen
  \bibfield  {author} {\bibinfo {author} {\bibfnamefont {M.}~\bibnamefont
  {Deniz}} \emph {et~al.} (\bibinfo {collaboration} {TEXONO}),\ }\href
  {\doibase 10.1103/PhysRevD.82.033004} {\bibfield  {journal} {\bibinfo
  {journal} {Phys. Rev.}\ }\textbf {\bibinfo {volume} {D82}},\ \bibinfo {pages}
  {033004} (\bibinfo {year} {2010}{\natexlab{b}})},\ \Eprint
  {http://arxiv.org/abs/1006.1947} {arXiv:1006.1947 [hep-ph]} \BibitemShut
  {NoStop}%
\bibitem [{\citenamefont {Xu}(2019)}]{Xu:2019dxe}%
  \BibitemOpen
  \bibfield  {author} {\bibinfo {author} {\bibfnamefont {X.-J.}\ \bibnamefont
  {Xu}},\ }\href {\doibase 10.1103/PhysRevD.99.075003} {\bibfield  {journal}
  {\bibinfo  {journal} {Phys. Rev.}\ }\textbf {\bibinfo {volume} {D99}},\
  \bibinfo {pages} {075003} (\bibinfo {year} {2019})},\ \Eprint
  {http://arxiv.org/abs/1901.00482} {arXiv:1901.00482 [hep-ph]} \BibitemShut
  {NoStop}%
\bibitem [{\citenamefont {Rosen}(1982)}]{Rosen:1982pj}%
  \BibitemOpen
  \bibfield  {author} {\bibinfo {author} {\bibfnamefont {S.~P.}\ \bibnamefont
  {Rosen}},\ }\href {\doibase 10.1103/PhysRevLett.48.842} {\bibfield  {journal}
  {\bibinfo  {journal} {Phys. Rev. Lett.}\ }\textbf {\bibinfo {volume} {48}},\
  \bibinfo {pages} {842} (\bibinfo {year} {1982})}\BibitemShut {NoStop}%
\bibitem [{\citenamefont {Bilmis}\ \emph {et~al.}(2015)\citenamefont {Bilmis},
  \citenamefont {Turan}, \citenamefont {Aliev}, \citenamefont {Deniz},
  \citenamefont {Singh},\ and\ \citenamefont {Wong}}]{Bilmis:2015lja}%
  \BibitemOpen
  \bibfield  {author} {\bibinfo {author} {\bibfnamefont {S.}~\bibnamefont
  {Bilmis}}, \bibinfo {author} {\bibfnamefont {I.}~\bibnamefont {Turan}},
  \bibinfo {author} {\bibfnamefont {T.~M.}\ \bibnamefont {Aliev}}, \bibinfo
  {author} {\bibfnamefont {M.}~\bibnamefont {Deniz}}, \bibinfo {author}
  {\bibfnamefont {L.}~\bibnamefont {Singh}}, \ and\ \bibinfo {author}
  {\bibfnamefont {H.~T.}\ \bibnamefont {Wong}},\ }\href {\doibase
  10.1103/PhysRevD.92.033009} {\bibfield  {journal} {\bibinfo  {journal} {Phys.
  Rev.}\ }\textbf {\bibinfo {volume} {D92}},\ \bibinfo {pages} {033009}
  (\bibinfo {year} {2015})},\ \Eprint {http://arxiv.org/abs/1502.07763}
  {arXiv:1502.07763 [hep-ph]} \BibitemShut {NoStop}%
\bibitem [{\citenamefont {Dutta}\ \emph {et~al.}(2016)\citenamefont {Dutta},
  \citenamefont {Mahapatra}, \citenamefont {Strigari},\ and\ \citenamefont
  {Walker}}]{Dutta:2015vwa}%
  \BibitemOpen
  \bibfield  {author} {\bibinfo {author} {\bibfnamefont {B.}~\bibnamefont
  {Dutta}}, \bibinfo {author} {\bibfnamefont {R.}~\bibnamefont {Mahapatra}},
  \bibinfo {author} {\bibfnamefont {L.~E.}\ \bibnamefont {Strigari}}, \ and\
  \bibinfo {author} {\bibfnamefont {J.~W.}\ \bibnamefont {Walker}},\ }\href
  {\doibase 10.1103/PhysRevD.93.013015} {\bibfield  {journal} {\bibinfo
  {journal} {Phys. Rev.}\ }\textbf {\bibinfo {volume} {D93}},\ \bibinfo {pages}
  {013015} (\bibinfo {year} {2016})},\ \Eprint
  {http://arxiv.org/abs/1508.07981} {arXiv:1508.07981 [hep-ph]} \BibitemShut
  {NoStop}%
\bibitem [{\citenamefont {Dent}\ \emph {et~al.}(2017)\citenamefont {Dent},
  \citenamefont {Dutta}, \citenamefont {Liao}, \citenamefont {Newstead},
  \citenamefont {Strigari},\ and\ \citenamefont {Walker}}]{Dent:2016wcr}%
  \BibitemOpen
  \bibfield  {author} {\bibinfo {author} {\bibfnamefont {J.~B.}\ \bibnamefont
  {Dent}}, \bibinfo {author} {\bibfnamefont {B.}~\bibnamefont {Dutta}},
  \bibinfo {author} {\bibfnamefont {S.}~\bibnamefont {Liao}}, \bibinfo {author}
  {\bibfnamefont {J.~L.}\ \bibnamefont {Newstead}}, \bibinfo {author}
  {\bibfnamefont {L.~E.}\ \bibnamefont {Strigari}}, \ and\ \bibinfo {author}
  {\bibfnamefont {J.~W.}\ \bibnamefont {Walker}},\ }\href {\doibase
  10.1103/PhysRevD.96.095007} {\bibfield  {journal} {\bibinfo  {journal} {Phys.
  Rev.}\ }\textbf {\bibinfo {volume} {D96}},\ \bibinfo {pages} {095007}
  (\bibinfo {year} {2017})},\ \Eprint {http://arxiv.org/abs/1612.06350}
  {arXiv:1612.06350 [hep-ph]} \BibitemShut {NoStop}%
\bibitem [{\citenamefont {Cui}\ \emph {et~al.}(2018)\citenamefont {Cui},
  \citenamefont {Pospelov},\ and\ \citenamefont {Pradler}}]{Cui:2017ytb}%
  \BibitemOpen
  \bibfield  {author} {\bibinfo {author} {\bibfnamefont {Y.}~\bibnamefont
  {Cui}}, \bibinfo {author} {\bibfnamefont {M.}~\bibnamefont {Pospelov}}, \
  and\ \bibinfo {author} {\bibfnamefont {J.}~\bibnamefont {Pradler}},\ }\href
  {\doibase 10.1103/PhysRevD.97.103004} {\bibfield  {journal} {\bibinfo
  {journal} {Phys. Rev.}\ }\textbf {\bibinfo {volume} {D97}},\ \bibinfo {pages}
  {103004} (\bibinfo {year} {2018})},\ \Eprint
  {http://arxiv.org/abs/1711.04531} {arXiv:1711.04531 [hep-ph]} \BibitemShut
  {NoStop}%
\bibitem [{\citenamefont {Abdullah}\ \emph {et~al.}(2018)\citenamefont
  {Abdullah}, \citenamefont {Dent}, \citenamefont {Dutta}, \citenamefont
  {Kane}, \citenamefont {Liao},\ and\ \citenamefont
  {Strigari}}]{Abdullah:2018ykz}%
  \BibitemOpen
  \bibfield  {author} {\bibinfo {author} {\bibfnamefont {M.}~\bibnamefont
  {Abdullah}}, \bibinfo {author} {\bibfnamefont {J.~B.}\ \bibnamefont {Dent}},
  \bibinfo {author} {\bibfnamefont {B.}~\bibnamefont {Dutta}}, \bibinfo
  {author} {\bibfnamefont {G.~L.}\ \bibnamefont {Kane}}, \bibinfo {author}
  {\bibfnamefont {S.}~\bibnamefont {Liao}}, \ and\ \bibinfo {author}
  {\bibfnamefont {L.~E.}\ \bibnamefont {Strigari}},\ }\href {\doibase
  10.1103/PhysRevD.98.015005} {\bibfield  {journal} {\bibinfo  {journal} {Phys.
  Rev.}\ }\textbf {\bibinfo {volume} {D98}},\ \bibinfo {pages} {015005}
  (\bibinfo {year} {2018})},\ \Eprint {http://arxiv.org/abs/1803.01224}
  {arXiv:1803.01224 [hep-ph]} \BibitemShut {NoStop}%
\end{thebibliography}%

\end{document}